\documentclass{aastex}
\usepackage{emulateapj5}
\usepackage{onecolfloat5}
\usepackage{natbib}
\bibpunct{(}{)}{;}{a}{,}{,}

\input epsf

\def\plotthree#1#2#3{\centering \leavevmode
    \epsfxsize=0.65\columnwidth \epsfbox{#1} \hfil
    \epsfxsize=0.65\columnwidth \epsfbox{#2} \hfil
    \epsfxsize=0.65\columnwidth \epsfbox{#3}}



\long\def\comment#1{}

\def\W2{{\cal W}}

\def\be{\begin{equation}}
\def\ee{\end{equation}}
\def\bea{\begin{eqnarray}}
\def\eea{\end{eqnarray}}

\def\cmm2{{\,\rm cm^{-2}}}
\def\cm2{{\,{\rm cm}^2}}
\def\cmm3{{\,{\rm cm}^{-3}}}
\def\gcmm3{{\,{\rm g\,cm^{-3}}}}

\def\fun#1#2{\lower3.6pt\vbox{\baselineskip0pt\lineskip.9pt
  \ialign{$\mathsurround=0pt#1\hfil##\hfil$\crcr#2\crcr\sim\crcr}}}

\lefthead{Dick et al.} \righthead{Reduction of Cosmological Data for the Detection of Time-varying Dark Energy Density}

\begin{document}
\bibliographystyle{apj}
\twocolumn[
\submitted{Submitted to JCAP}
\title{Reduction of Cosmological Data for the Detection of Time-varying Dark Energy Density}
\author{Jason Dick$^1$, Lloyd Knox$^1$ and Mike Chu$^2$}
\affil{$^{1}$ Department of Physics, University of California,
Davis, CA 95616, USA, email:
jadick@ucdavis.edu,lknox@ucdavis.edu,chu@milkyway.gsfc.nasa.gov}
\affil{$^2$ Goddard Space Flight Center, Greenbelt, MD 20771}

\begin{abstract}
We present a method for reducing cosmological data to constraints on
the amplitudes of modes of the dark energy density as a function of
redshift.  The modes are chosen so that 1) one of them has constant
density and 2) the others are non-zero only if there is time-variation
in the dark energy density and 3) the amplitude errors for the
time-varying modes are uncorrelated with each other.  We apply our
method to various combinations of three-year WMAP data \citep{spergel06}, 
baryon acoustic
oscillation data \citep{eisenstein05}, the \citet{riess04} 'Gold'
supernova data set, and the Supernova Legacy Survey data set
\citep{astier05}.  We find no significant evidence for a time-varying
dark energy density or for non-zero mean curvature.  Although by
some measure the limits on four of the time-varying mode amplitudes 
are quite tight, they are consistent with the expectation that
the dark energy density does not vary on time scales shorter than a 
Hubble time.  Since we do not expect detectable time variation in these
modes, our results should be viewed as a systematic error test which
the data have passed.  We discuss a procedure to identify modes with maximal
signal-to-noise ratio. 
\end{abstract}

\keywords{cosmology: theory -- cosmology: observation} ]

\section{Introduction}

That the cosmological expansion rate is accelerating is 
well-established\footnote{See for
example \citet{shapiro05} and references therein.}.  Two pressing
questions, of deep importance for fundamental physics, are 1) Is the
acceleration
due to corrections to the gravitational field equations or to some
unknown matter component that we call dark energy? and 2) If dark
energy, is it a cosmological constant or something with a time-varying
density?  Here we present a method for reducing cosmological
measurements of distance as a function of redshift in a manner suited
to answering this second question.

Our method reduces cosmological data (anything that depends on the
history of the dark energy density, $\rho_x(z)$) to constraints on a
cosmological constant plus the amplitudes of time-varying modes of
$\rho_x(z)$.  The modes are chosen to have uncorrelated amplitude
errors and to be those that are best determined by the data.  The two
chief desirable properties of such a reduction are:
\begin{enumerate}
\item The cosmological constraints can be expressed in a highly
model-independent manner in terms of just a few numbers (plus
associated functions of redshift).
\item Consistency with a cosmological constant is straightforward
to study visually from the graphed results.
\end{enumerate}

Many have previously considered different methods for parameterizing
possible departures from a cosmological constant
\citep{chevallier01,weller02,linder03b,huterer03}.
Such parameterizations, while not necessary for comparing the relative
merits of two dark energy models, facilitate interpretation of the
data in a less model-dependent manner.  Most of these
are one or two-dimensional parameterizations of $w(z)$.

The most frequently used parameterizations are $w(z)$ is constant and
$w(z) = w_0 + \left(1-a\right)w_a$ where $a = 1/(1+z)$ is the scale
factor \citep{chevallier01,linder03b}.  However, even allowing for
non-zero $w_a$, these allow only for departures from constant density
in a highly restricted space among the space of all possible ways the
density could vary.  It is possible for experiments to be sensitive to
time variation and yet have $w_0 =-1$ and $w_a = 0$ to within the
uncertainties.  Additionally, current data do not constrain $w_0$ and
$w_a$ very well, and \citet{simpson06} find that the parameterization
can lead to significant biases in the inferred $w(z)$ and other
cosmological parameters.

We have chosen to work with density because it has
a more direct relation to the data.  The data we consider depend
on distance as a function of redshift.  This function can be calculated
from $\rho_x(z)$ with one integral while calculating it from $w(z)$ requires
two integrals.  

\citet{wang06}, \citet{daly04} and \citet{wang04} reconstruct
$\rho_x(z)$ from distance and other cosmological data. Although these
reconstructions are useful for visual inspection, detailed
interpretation of results is obscured by the correlation of errors
across redshift.  To ameliorate this difficulty, \citet{huterer05}
reduce data to weighted averages of $\rho_x(z)$ with uncorrelated
errors.  This has advantages, but the drawback that to look for
time-variation one must visually differentiate the data.

Although we are working with $\rho_x(z)$, we are not attempting a
reconstruction of this function from the data.  We identify the
best-determined time-varying modes of $\rho_x(z)$ and then determine
the probability distribution of their amplitudes, having marginalized
over all the other parameters, including the amplitude of the constant
mode.  If any of these mode amplitudes is significantly non-zero, we
have evidence for time-varying dark energy.  By design, the amplitudes
of these errors are uncorrelated for ease of interpretation.

Somewhat similar to our approach has been the study of eigenmodes
of $w(z)$ \citep{huterer03}.  The eigenvalue spectra 
\citep{song04,knox04} suggest that $w(z)$ may eventually need to be described
with more than just two numbers, although for another view see
\citet{linder05a}.

Finally, we should mention that
\citet{wang05} reduce data to $H(z)$ in redshift bins with the
attractive property that the errors in $H(z)$ are uncorrelated from
bin to bin and only dependent on the supernovae in that bin.  However,
this reduction is not ideal for detecting a departure
from a cosmological constant, especially given uncertainty in
$\Omega_m$ and $\Omega_K$ which affect the conversion of $H(z)$ to
$\rho_x(z)$.  

Our method is applicable to any measurements and
any model space in which dark energy's sole influence comes
through the history of its energy density, $\rho_x(z)$.  We apply
it here to two different supernova data sets (the Gold \citep{riess04} 
and SNLS \citep{astier05}) with and without the baryon acoustic oscillation
(BAO) distance constraints from \citet{eisenstein05} and always
with constraints from CMB data.  The CMB data are important for how
it constrains the distance to last scattering and the matter density.

Our method may be even more attractive as descriptions of the
uncertainties in luminosity distances become more complex (due to the
increased importance of systematic relative to statistical error).  In
this case the use of luminosity distances as a final stage of data
reduction becomes more cumbersome.  A reduction to the constraints on
a few mode amplitudes may then be significantly easier to use than the
reduction to constraints on hundreds or thousands of luminosity
distances.

Identifying well-determined modes is useful not only for detection of
time-variation, but also for ferreting out systematic errors.  With
the best-determined modes, one can split the data up into subsets and
check the corresponding mode amplitude estimates for consistency.
Since the amplitudes have small errors, they can survive the greatest
amount of sub-sampling while remaining sufficiently well-measured
that the consistency tests remain meaningful.

We compare the data-determined limits on the mode amplitudes to
theoretical upper limits derived from the criterion that the
time-scale for dark energy density variation is longer than a Hubble
time.  We find these Hubble time upper limits to be tighter than the
data-determined upper limits for all modes.  Thus, any detection of
non-zero mode amplitude with current data would most likely be an
indication of systematic error.  Due to this theoretical expectation
of a null result we conclude that our method, as currently implemented
on current data, is best though of as a powerful method for
identifying otherwise undetected systematic errors.  We discuss
means of folding in theoretical expectations for the level of
time variation so that we can identify modes that maximize
signal-to-noise ratio, rather than just minimize noise.

In our applications we make no assumption about the mean
curvature.  We allow it to vary, constrained only by the data, and
plot the result together with $\Omega_x$ and the amplitude of the
time-varying dark energy density modes.  We do this to avoid what
would be the highly unfortunate mistake of declaring detection of time
variation when the data could just as well be explained by non-zero
mean curvature.  \citet{linder05d} has recently demonstrated the
extent of the $\Omega_K-w_0-w_a$ degeneracy for future CMB + supernova data,
and the degeneracy between $\Omega_K$ and $w_0$ for the SDSS BAO
data is commented on in \citet{eisenstein05}.  We also see
determination of the (possibly non-zero) mean curvature as another
very interesting application of distance vs. redshift measurements
\citep{bernstein05,knox06,freivogel05}.

In section II we describe the method in detail.  In section III
we apply it to the SNLS + CMB data.  Results from
other combinations of supernova, CMB and BAO data are presented in
Appendix B.  In section IV we discuss our results and conclude.  

\section{Method}
The goal of our analysis is to demonstrate a method for measuring
non-constant dark energy with a combination of low-$z$ distance
measurements, CMB data, and BAO constraints.  In
\ref{sec:param} we describe our parameterization of the cosmology.  
In \ref{sec:likelihood} we describe how the likelihood of these parameters
is calculated given each of the data sets.  In \ref{sec:diag} we describe
our calculation of the eigenmodes.  In \ref{sec:mcmc} we describe our use
of the Monte Carlo Markov Chain method that we use to estimate the parameters
and their uncertainties.

\subsection{Parametrization}
\label{sec:param}

\begin{table}
\caption{Cosmological Parameters}
\begin{center}
\begin{tabular}{ll}
\hline
\\
$\omega_m$ & Matter density: $\omega_m \equiv \Omega_mh^2$, $h = \frac{H_0}{100 {\rm km/sec/Mpc}}$ \\
$\Omega_k$ & Curvature: $\Omega_k \equiv -k\frac{c^2}{H_0^2} = 1-\Omega_{tot}$ \\
$\alpha_i$ & Dark energy parameters: $\rho_x(z) = \rho_c\sum_i\alpha_i e_i(z) $ \\
\\
\hline
\label{tbl:params}
\end{tabular}
\end{center}
\end{table}

Our set of cosmological parameters is the matter density today,
$\omega_m$ in units of 1.88 $\times 10^{-29}$ g/cm$^3$, $\Omega_k
\equiv -k c^2/H_0^2 = 1-\Omega_{\rm tot}$ and the $\alpha_i$ that
determine the history of the dark energy density $\rho_x(z)$.  Given
the basis functions, $e_i(z)$ (to be described in \ref{sec:diag}), the
dimensionless parameters $\alpha_i$ specify $\rho_x(z)$ up to an
overall constant which is the critical density today, $\rho_c$.  These
cosmological parameters are summarized in Table \ref{tbl:params}.

The supernova data sets also include some `nuisance' parameters
required to model the data because of our inability to infer precisely
the luminosity of each of the supernovae.  These parameters are
described in the appropriate likelihood calculation subsections
(\ref{sec:riesslike} and \ref{sec:snlslike}).

\subsection{Likelihood Calculation}
\label{sec:likelihood}
Our constraints come from combinations of WMAP3 data
\citep{spergel06}, a BAO constraint from
\citet{eisenstein05}, and supernova data from \citet{riess04} and
\citet{astier05}.  Taking uniform priors on the parameters, the
probability distributions of the parameters given the data are
proportional to the likelihood functions, ${\cal L}$.  Since we make
no use of the probability amplitudes here, we use likelihood and
probability interchangeably in what follows.

\subsubsection{CMB Likelihood}
\label{sec:CMBL}
CMB data are sensitive to a large number of cosmological parameters,
but only two of them are relevant for interpreting data that is
sensitive to the expansion rate at lower redshifts, such as supernova
data and BAO data.  These are 
$\omega_m$ and the comoving angular-diameter distance to 
last scattering, $D_{\ast} = D_M(z_*)$
where 
\be
D_M(z) = \frac{1}{\sqrt{k}} \mathrm{sin} [ c \sqrt{k} \int_0^z \frac{dz'}{H(z')}]
\label{eqn:D_M}
\ee
and $z_*$ is the redshift of the surface of last scattering, here taken
as the peak of the visibility function.  We therefore derive, from
the CMB data, a likelihood distribution for the two-dimensional parameter
space:  $\omega_m, D_\ast$.  

To calculate this likelihood function we use an MCMC chain calculated
from the WMAP3 data set and available on the LAMBDA
archive\footnote{Legacy Archive for Microwave Background Data
Analysis: http://lambda.gsfc.nasa.gov}.  We use a chain which uses the
simplest-case $\Lambda$CDM model with only WMAP data, which assumes
zero curvature and a cosmological constant.  However, at fixed
$\omega_m$ and $D_\ast$ the CMB data are highly insensitive to
departures from a cosmological constant or non-zero curvature.  A
chain that allowed for curvature and dark energy would not give a
significantly different likelihood function for $\omega_m$ and $D_*$
as long as the dark energy and curvature remained sub-dominant at last
scattering.  The dominant effect of curvature or dark energy on the
CMB is to change the projection of length scales on the
last-scattering surface to observed angular scales because of the
influence of $k$ and $H(z)$ on $D_M(z_*)$ in Eq.~\ref{eqn:D_M}.  The
constraint from CMB data on $D_M(z_*)$ thus captures the CMB
information about $\Omega_k$ and dark energy.

Prior to the recent WMAP3 release we calculated the likelihood of
$\Omega_m h^2$ and $D_*$ from an MCMC chain calculated in
\citet{chu05} from WMAP1, Cosmic Background Imager (CBI) and Arcminute
Cosmology Bolometer Array Receiver (ACBAR) data -- the `WMAPext'
data set used in \citet{spergel03}.  Our update to WMAP3 does
not have a significant impact on our dark energy results.

There is some weak information from the CMB about dark energy and
curvature, beyond that in $\omega_m$ and $D_*$, coming from the
integrated Sachs Wolfe (ISW) effect 
\citep[for a review of CMB physics see][]{hu02e}.  Ideally 
our CMB likelihood
function would capture this information.  For simplicity we ignore it,
because the constraints are very weak and model-dependent. 

To calculate our CMB likelihood function, we begin by calculating the
parameters $D_*$ and $\omega_m$ at every chain step to create a new
chain in these two parameters.  By counting the number of chain
elements in each bin of a regular grid in $\omega_m$ and $D_\ast$ we
create a two-dimensional matrix that is an approximation to the
probability distribution $P_{CMB} ( \omega_m, D_{\ast} )$ given by the
WMAP3 data. Due to the properties of MCMC, the number of chain
elements that fall within each bin will be proportional to the
probability in that bin plus sample variance fluctuations.  Since we
have ignored the other chain parameters in this process, such as the
optical depth to last scattering and the primordial power spectrum
spectral index, we have, in effect, marginalized over them.

Before making use of this matrix, we attempt to reduce the noise
caused by the finite number of samples by implementing a low-pass
filter.  We do so by taking a Fast Fourier Transform of the two
dimensional matrix above and eliminating the high frequency
components.  The inverse Fourier transform of this new matrix is a
noise-reduced approximation to the probability distribution. An
appropriate frequency cut is chosen with two criteria: the difference
between the original matrix and the noise-reduced matrix can be
well-explained by Poisson noise, and the majority of the noise is
eliminated (that is, neighboring bins do not vary dramatically).

\begin{figure}[t]
\centerline{\scalebox{.4}{\includegraphics{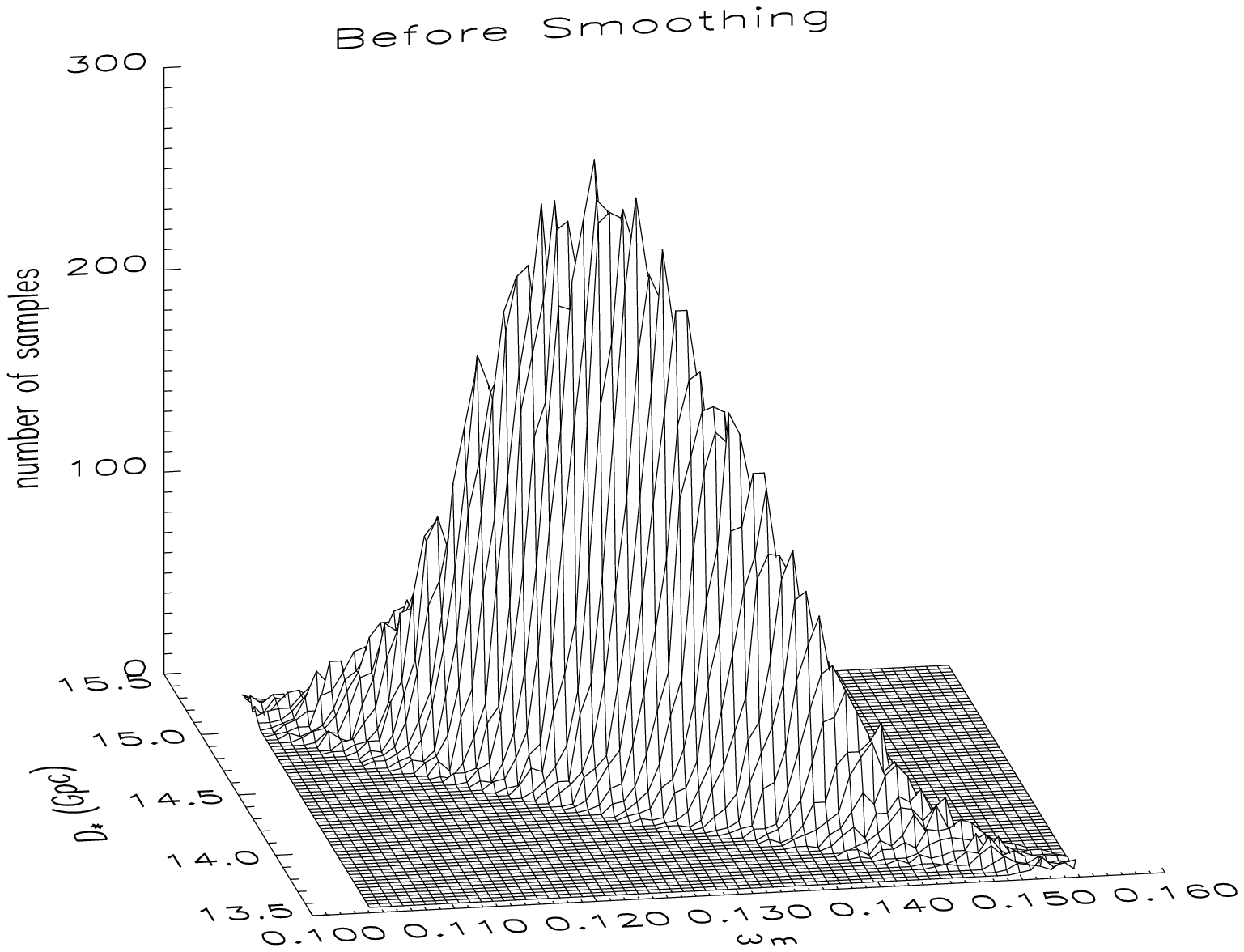}}}
\centerline{\scalebox{.4}{\includegraphics{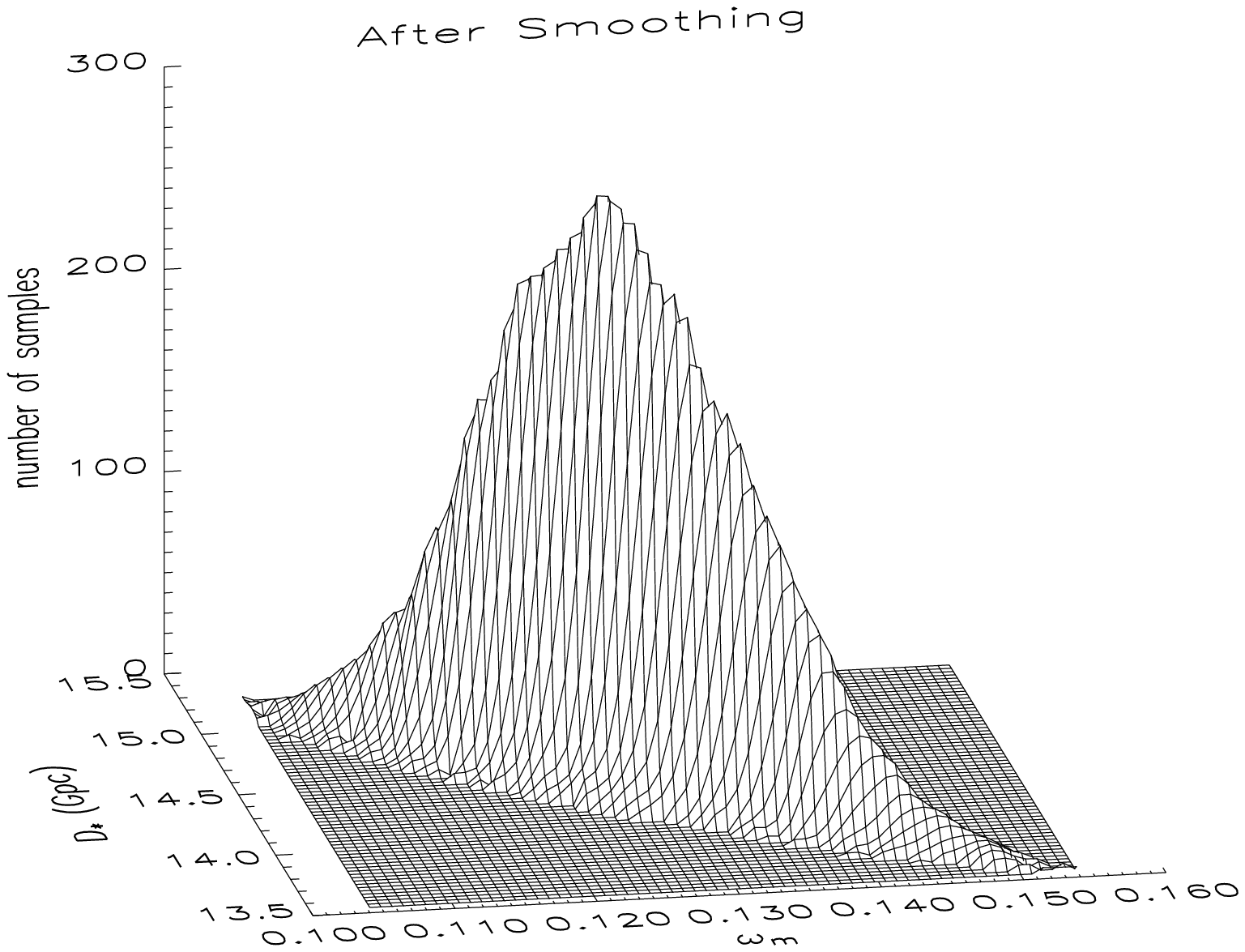}}}
\caption{Probability distribution of $\omega_m$ and $D_*$ given
CMB data before smoothing (top) and after smoothing (bottom).  Note that
the large-scale features are preserved, while the small-scale,
sample-variance-induced noise has been greatly reduced.}
\label{fig:smooth}
\end{figure}

Due to the inaccuracy of both the original chain and this
noise-reduced matrix for very low probabilities, we set to zero all
bins which include fewer chain steps than some cutoff value, with this
cutoff value chosen so that fewer than 0.003 times the total number of
chain elements are excluded when these bins are set to zero.  If this
zeroing is not done, then the low probability regions of the
noise-reduced matrix are dominated by ringing and include negative
values.  The result of reducing the noise on an example probability
distribution $P_{CMB}(\omega_m, D_*)$ is shown in
Fig.~\ref{fig:smooth}.  This noise-reducing algorithm allows us to
have an approximation to the full chain that is closer to the chain
output than a Gaussian approximation, has no obvious noise, and does
not have bins that are so coarse that detail is lost.  We evaluate
$P_{CMB}(\omega_m, D_{\ast})$ by bilinear interpolation
over the smoothed grid.

We have not included some other recent CMB results, most importantly
the 2003 flight of Boomerang \citep{montroy05,piacentini05,jones05}
and the most recent results from CBI \citep{readhead04a,readhead04b}. 
Including these would further tighten up the CMB constraints
on $\omega_m$ and $D_{\ast}$ \citep{mactavish05,sievers05}.  Although,
given the lack of change from WMAPext to WMAP3, we do not expect
that these additional data would be significant for our application.

\subsubsection{BAO Likelihood}

The next step is to incorporate the BAO data from luminous red
galaxies in the SDSS survey.  For this we use the parameter $D_V$
given in \citet{eisenstein05} equation (2):

\be D_V(z) \equiv [D_M^2(z) \frac{cz}{H(z)}]^{1/3} \mathrm{.} \ee

\citet{eisenstein05} have compressed the full data set into a function
of $D_V$ in two separate ways.  For this paper, we select the $A$
parameterization, their equation (4), because it
appears to give tighter constraints on the dark energy:
\be 
A \equiv D_V(0.35)\frac{\sqrt{\Omega_mH_0^2}}{0.35c} = 0.469 \pm 0.017(3.6\%)\mathrm{.}
\ee
Though this BAO data is a very powerful constraint, it is unfortunate
for our analysis that it was reduced to a constraint at a single
redshift.  This reduction was intended to be valid for the case of a
cosmological constant, and fairly robust against changes in constant
$w(z)$, but may introduce significant systematic errors for the cases
we consider with much more freedom in the behavior of the dark energy.
It would be much more useful to have the data reduced to distance
constraints for multiple redshift bins, and we encourage such
reductions from future analyses of BAO data.

We write $\chi^2 \equiv -2\ln{\cal L}$ for the BAO data set as:
\be \chi_{BAO}^2 = \frac{(A - 0.469)^2}{0.017^2} \mathrm{.} \ee

\subsubsection{Gold Likelihood}
\label{sec:riesslike}
The observed magnitude $m$ of an object can be linked to cosmology by:
\be
m = M +5\mathrm{log}_{10}\left(\frac{D_L(z)}{10\mathrm{pc}}\right)\mathrm{,}
\label{eq:m}
\ee
where the absolute magnitude $M$ is defined as the hypothetical
observed magnitude of an identical object at $10\mathrm{pc}$, and the
luminosity distance $D_L(z)$ is evaluated at the redshift of the
object in question.  The luminosity distance is simply-related to the
comoving angular-diameter distance defined in equation \ref{eqn:D_M} by
a factor of $(1+z)$:

\be
D_L(z) = (1+z)D_M(z)\mathrm{.}
\ee

  We can then define the distance modulus $\mu$ as the portion of
equation \ref{eq:m} that is dependent only upon the cosmology as:

\be
\mu = m - M = 5\mathrm{log}_{10}\left(\frac{D_L\left(z\right)}{10\mathrm{pc}}\right)\mathrm{.}
\ee

For the Gold data set, the distance modulus $\mu$ is estimated
for each supernova.  But since the absolute magnitude $M$ is unknown,
we consider the $\mu$ for each supernova to be:
\be
\mu_i(\delta M) = \mu^d_i - \delta M\mathrm{,}
\ee
where $\mu^d_i$ is given for each supernova in the Gold data set from
\citet{riess04}.  The $\mu^d_i$ have been estimated by Riess
et. al. from observations in order to correct the observed magnitude
of each supernova for dust, as well as differences in magnitude
between supernovae that are correlated with the shape of the
luminosity versus time function.  In doing this estimation to generate
the Gold data set, a mean value of the absolute magnitude $M$ has been
assumed.  The parameter $\delta M$ is the difference between the true
mean absolute magnitude of the supernovae and the estimated absolute
magnitude.  This parameter is marginalized over.  To evaluate $\chi^2$
for a cosmology with a distance-redshift relation $D_L(z)$ we
calculate:

\be
\chi^2 = \sum_i \frac{\left( \mu_i(\delta M)  - 5 \mathrm{log}_{10} \left( \frac{D_L\left(z_i\right)}{10\mathrm{pc}}\right) \right)^2}
{\sigma_i^2}\mathrm{.}
\ee

This assumes Gaussian noise in the distance moduli quoted in the Gold
data set, and the sum over $i$ is over the various supernovae.

\subsubsection{SNLS Likelihood}
\label{sec:snlslike}
The SNLS data are reduced differently, providing us with constraints
on the parameters used to calculate the effective apparent magnitude
rather than just the effective apparent magnitude itself.  For each
supernova, the stretch factor $s$, color factor, $c$, and rest frame B
band apparent magnitude $m_B^*$ are estimated from the light curves.
We can then define a distance modulus $\mu$ for each supernova which
can be directly compared to the cosmology:

\be
\mu_i(M, \alpha, \beta) = m_B^* - M + \alpha (s_i-1) - \beta c_i.
\ee

For our likelihood calculation we take the best fit values of $m^*_B$,
$s$, and $c$ from \citet{astier05} for each supernova, and then treat
$M$, $\alpha$, and $\beta$ as global parameters free to vary just like
the cosmological parameters.

Since we do not have a covariance matrix for the parameters $m^*_B$,
$s$, and $c$, we simply fix these quantities to their best fit values
for each supernova and use the uncertainty in $\mu_B$ provided in
\citet{astier05} to form

\be
\chi^2 = \sum_i \frac{\left( \mu_i(M, \alpha, \beta) - 5 \mathrm{log}_{10} \left( \frac{D_L(z_i)}{10\mathrm{pc}}\right) \right)^2}{\sigma^2_i(\mu_B) +\sigma_{\rm int}^2}\mathrm{.}
\ee

Here $\sigma^2_{\rm int}$ is the intrinsic dispersion of the absolute
magnitudes.  We take the value of $\sigma_{\rm int} = 0.15$ for the
nearby supernovae, and $\sigma_{\rm int} = 0.12$ for the SNLS
supernovae, the mean values reported in \citet{astier05}.

Residuals of the binned SNLS distance modulus data, after subtraction
of the mean distance modulus for the $\Lambda$CDM model, are shown
in Fig.~\ref{fig:onesigvar}.  The binning is purely for plotting purposes.
For calculation of the likelihood function we use the unbinned data.

\subsubsection{Final Likelihood}

We now have three independent data sets: CMB, BAO, and one of two
supernova data sets.  Since they are independent, the probabilities
multiply:
\be
\mathrm{ln}({\cal L}_{\rm TOT}) = \mathrm{ln}(P_{CMB}(\omega_m, D_*)) - \frac{\chi^2_{SN}}{2} - \frac{\chi^2_{BAO}}{2},
\ee
where $\chi^2_{SN}$ comes from either the Gold or SNLS data set.
Because the SNLS and Gold data sets make use of many of the same
nearby supernovae, but use different algorithms for correcting the
magnitudes, the data sets are correlated.  Combining them
appropriately would require more than a simple summing of the Fisher
matrices and is an exercise we have not attempted.

\subsection{Eigenmode Calculation}
\label{sec:diag}
To select our modes
we need the covariance matrix for the errors in the $\alpha_i$
parameters for some initial basis. One could in principle
calculate the covariance matrix from an MCMC chain, 
but since a chain run with the full
set of $\alpha_i$ parameters is highly non-Gaussian, the calculated
covariance matrix is not a good approximation and does not give good
eigenmodes. Instead we estimate it analytically
from the Fisher matrix, and then use this estimate to define the modes.

\subsubsection{The Fisher Matrix}
First, we generate a Fisher matrix by expanding about a particular point. The
point chosen is the mean point of a chain run with the aforementioned
likelihood function, but using $\Omega_{\Lambda}$ instead of $\alpha_i$. This
Fisher matrix is calculated in two parts.  For the supernova part we obtain:

\be
F^{SN}_{ij} = \sum_k \frac{\partial x_k}{\partial p_i}\frac{1}{\sigma_k^2} \frac{\partial x_k}{\partial p_j}
\ee
where for SNLS $p_k = ( \omega_m, \Omega_k, M,
\alpha, \beta, \alpha_0, \alpha_1, \alpha_2, ..., \alpha_{100} )$ and
\be
x_k \equiv 5\mathrm{log}_{10}\left(\frac{D_L(z_k)}{10\mathrm{pc}}\right) + 
M - \alpha(s_k - 1) + \beta c_k
\ee
and for Gold $p_k = ( \omega_m, \Omega_k, \delta M,
\alpha_0, \alpha_1, \alpha_2, ..., \alpha_{100} )$ and
\be
x_k \equiv
5\mathrm{log}_{10}\left(\frac{D_L(z_k)}{10\mathrm{pc}}\right) + \delta M.
\ee

Since the BAO data are written as a single constraint in
$A$, which is a function of the chain parameters, calculating
its Fisher matrix is even simpler:
\be F^{BAO}_{ij} = \frac{\partial A}{\partial
p_i}\frac{1}{\sigma^2}\frac{\partial A}{\partial p_j}. \ee

Once this is calculated, we need to factor in the CMB data. Because
performing numerical derivatives on our two-parameter probability
matrix would be numerically unstable, we first calculate a covariance
matrix from the chain calculated in section \ref{sec:CMBL} which
includes only $\omega_m$ and $D_{\ast}$, then invert this covariance
matrix to obtain a Fisher matrix, which we will call $F^{2 x 2}$.  We
convert this Fisher matrix to one in the parameters of interest as
follows:
\be
F^{CMB}_{ij} = \left( \frac{\partial v}{\partial p_i} \right)^T F^{2 x 2} \left(\frac{\partial v}{\partial p_j} \right)
\ee
where $v$ is the vector defined by $v = \left[ \omega_m, D_{\ast} \right]$. 
Finally,
\be F_{ij}^{TOT} = F^{SN}_{ij} + F^{BAO}_{ij} + F^{CMB}_{ij}\mathrm{.} \ee
This total Fisher matrix is then inverted to give us a covariance matrix for
the eigenmode calculation.  The covariance matrix, once diagonalized,
gives us our eigenmodes.

\subsubsection{Initial Basis and Diagonalization}
\label{sec:initial}
Our specific goals for the desired modes are that there be one and
only one constant mode, and that the time-varying modes have
amplitudes with uncorrelated errors.  We can meet these two criteria
by careful selection of the initial basis and diagonalization
procedure.

The initial basis must have the following properties:
\begin{enumerate}
\item It must have one element which is a constant.
\item The elements must be linearly independent.
\end{enumerate}
Note that the second property is satisified by any basis by definition.
There are many possible bases that satisfy these criteria.  We have
chosen ours with the additional (somewhat arbitrary) criteria that
their dot products, $\sum_a e_i(z_a) e_j(z_a) = N \delta_{ij}$, where
$N$ is the dimensionality of the space given by the discretization
procedure.  The normalization factor, $N$, keeps the shapes and
amplitudes of the best-determined modes independent of $N$ as the
continuum limit is approached.  For completeness, we describe the
initial basis in Appendix A.

Each basis element is defined based on its value at $N$ values of $z$
between 0 and 2.  With the choice of uniform spacing in $\ln(0.01+z)$
we find we are able to approach the continuum limit with a smaller
value of $N$ than with other possible choices, such as uniform spacing
in $z$.  The results in this paper were generated with $N=50$.  Our
basis is piecewise-linear in $\ln(0.01+z)$.

From the procedure in Appendix A we get $N$ $\alpha_i$ parameters with
which to describe $\rho_x(z)$.  To obtain our eigenmodes we invert the
Fisher matrix in the parameter space of these $\alpha_i$ plus all our
other parameters to obtain a covariance matrix.  Then we take the $i>0$
subspace of this covariance matrix.  Diagonalizing this $N-1$ by $N-1$
covariance matrix gives us our non-constant eigenmodes, each of which
has errors uncorrelated with the other eigenmodes.  Including the
constant mode completes the space.  We call this new basis
$e^{\mathrm{em}}_i(z)$, and its coefficients $\alpha^{\mathrm{em}}_i$.

We typically truncate this new space by only keeping the first few
best-measured eigenmodes.  This results in an MCMC chain which is both
highly Gaussian and easy to interpret, since a significant departure
from zero of any one of the first few $\alpha^{em}_i$ parameters is a
signature of non-constant dark energy.  In the following we omit the
superscript ``em'' from the $\alpha_i$ parameters to reduce notational
clutter.

Note that the error in the constant mode has {\em not} been decorrelated
with the errors in the non-constant modes.  We discuss this choice
in Appendix A.

\subsection{Generating the Chain}
\label{sec:mcmc}
We use the Metropolis-Hastings algorithm to generate a Monte Carlo
Markov Chain \citep{gamerman97} as described in \citet{christensen01}.
Due to the low computational requirements we are able to run very long
chains with 6,010,000 elements.  We then ignore the first 10,000 steps
and thin the chain by taking every $20^{th}$ element, resulting in a
chain of length 300,000.

We can be sure that this thinned, 300,000-length chain has converged
by comparing the parameter estimates between different subsets of the
chain.  For example, if we take the subset of the first 5,000 elements
of the thinned chain with $6$ dark energy parameters for the SNLS +
CMB data set combination, and compare those parameter estimates to
those of the last 5,000 elements of the chain, those parameter
estimates usually vary by less than $\sigma/10$.  Depending on the
selection of elements, occasionally one of the 10 chain parameters
will vary by as much as $\sigma/5$.  Since the chain has converged to
an accuracy within about $\sigma/10$ at the 90\% confidence limit for
each parameter after 5,000 elements, and we run the chain to 300,000,
convergence is not an issue.

We use a generating function which is a multivariate normal distribution.  Its
covariance matrix is obtained from a pre-run.  This pre-run is done in two
iterations, each of length 100,000 with thinning by 20.  The second iteration
uses the covariance matrix from the first, with the covariance matrix from the
second giving the covariance matrix used to run the final chain.  This
iterative process gives us a generating function that is a good Gaussian
approximation to the full probability distribution.  This match between
generating function and posterior makes for an efficient exploration of 
the posterior.  

We set a (weak) prior constraints on $\Omega_m$ by bounding it
to the interval $0 < \Omega_m < 1$.

\section{Results}

\begin{figure}[b]
\centerline{\scalebox{.4}{\includegraphics{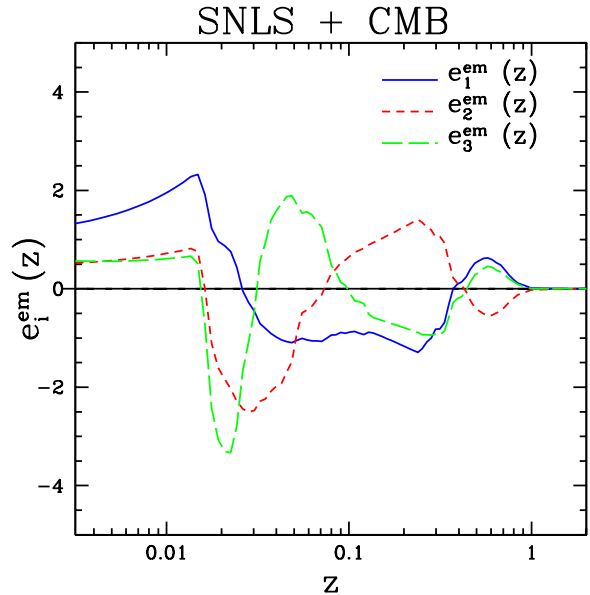}}}
\caption{The three best-measured eigenmodes for the SNLS + CMB data set.}
\label{fig:eigenmodes}
\end{figure}

We start by examining one data set combination in detail:  SNLS with
CMB data.  The first few eigenmodes are shown in figure \ref{fig:eigenmodes}.  
As is typical with such eigenmode decompositions, the frequency of oscillations 
tends to increase as the mode number increases.

\begin{figure*}[t]
\plotthree{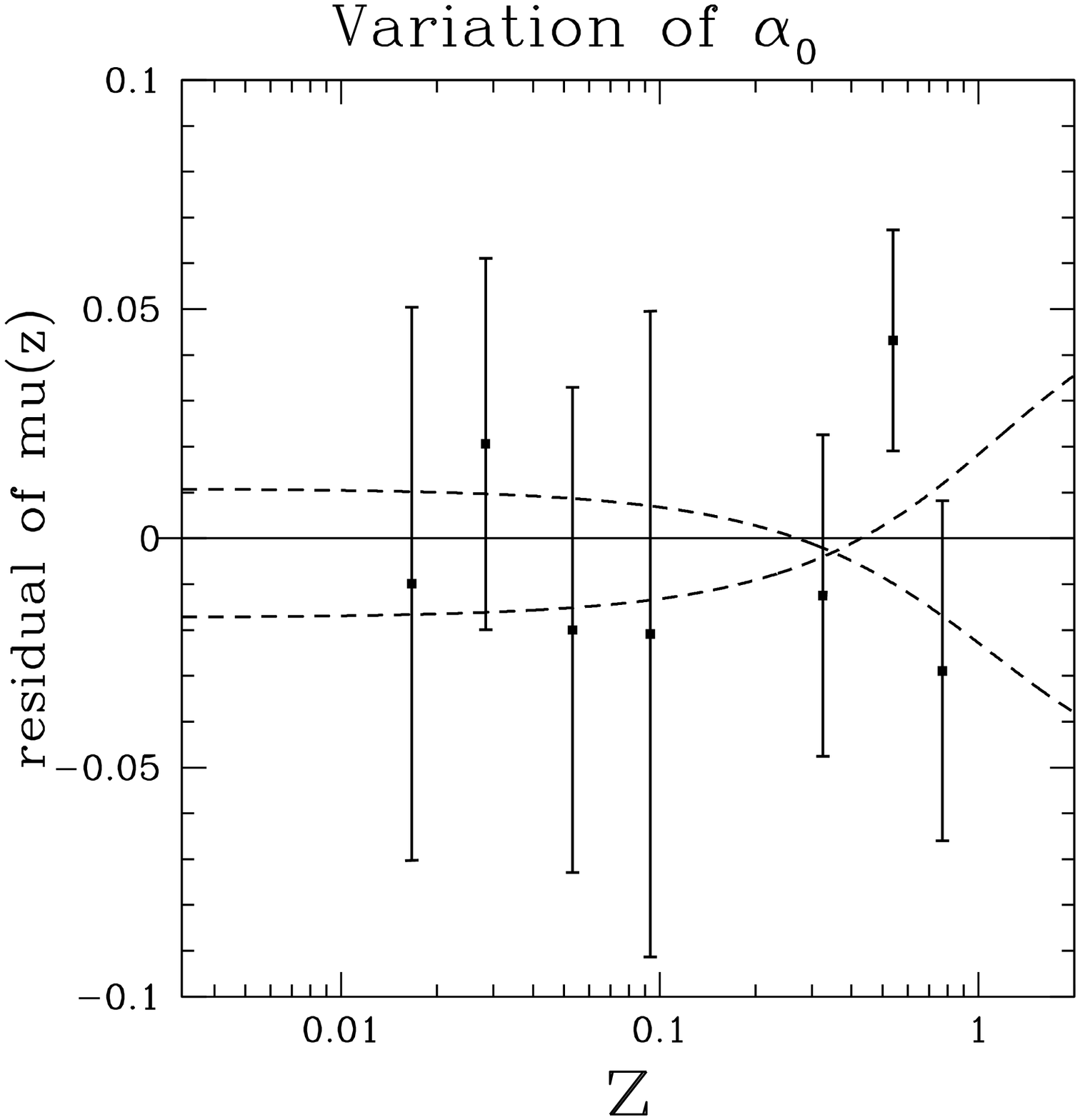}{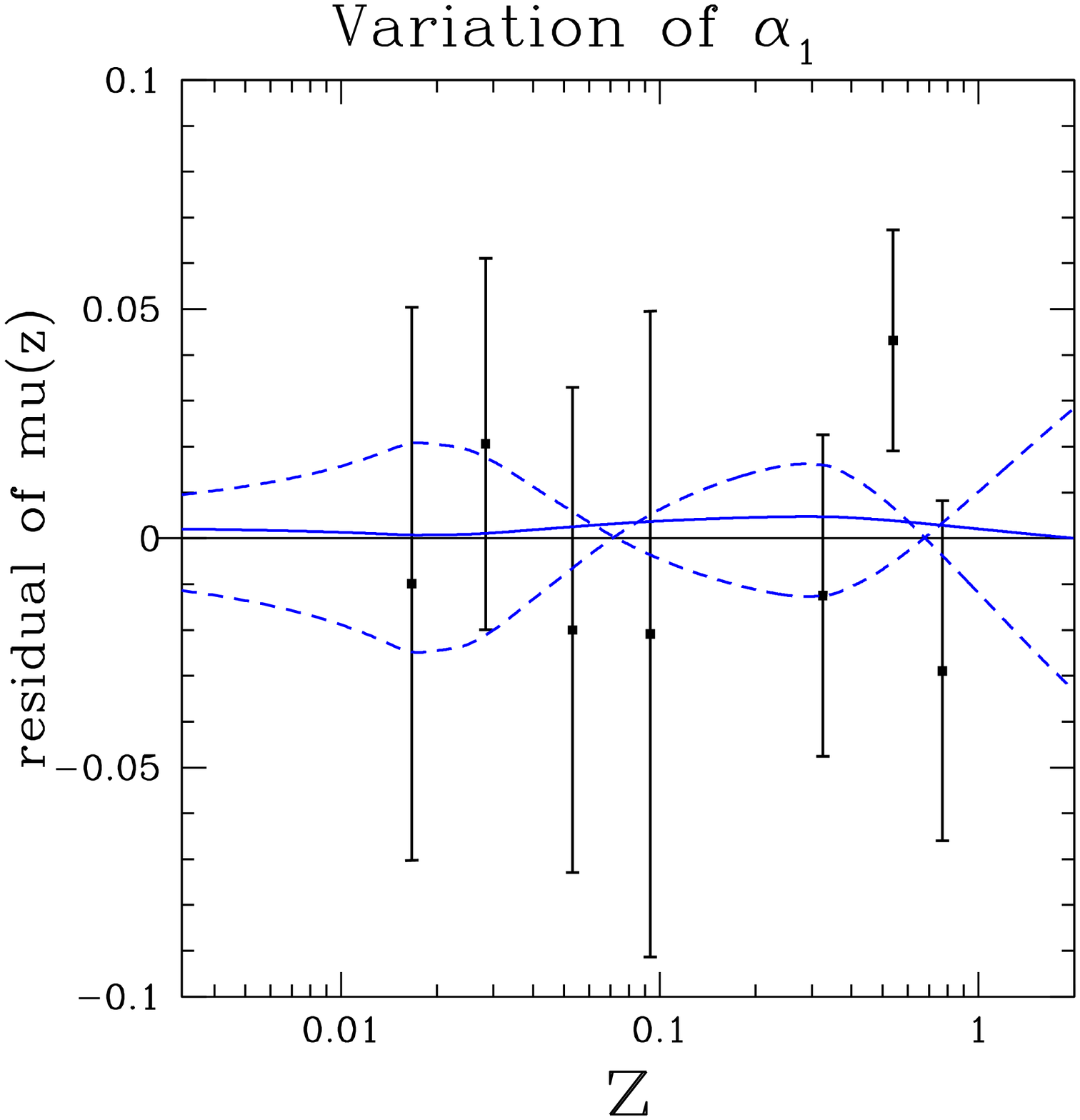}{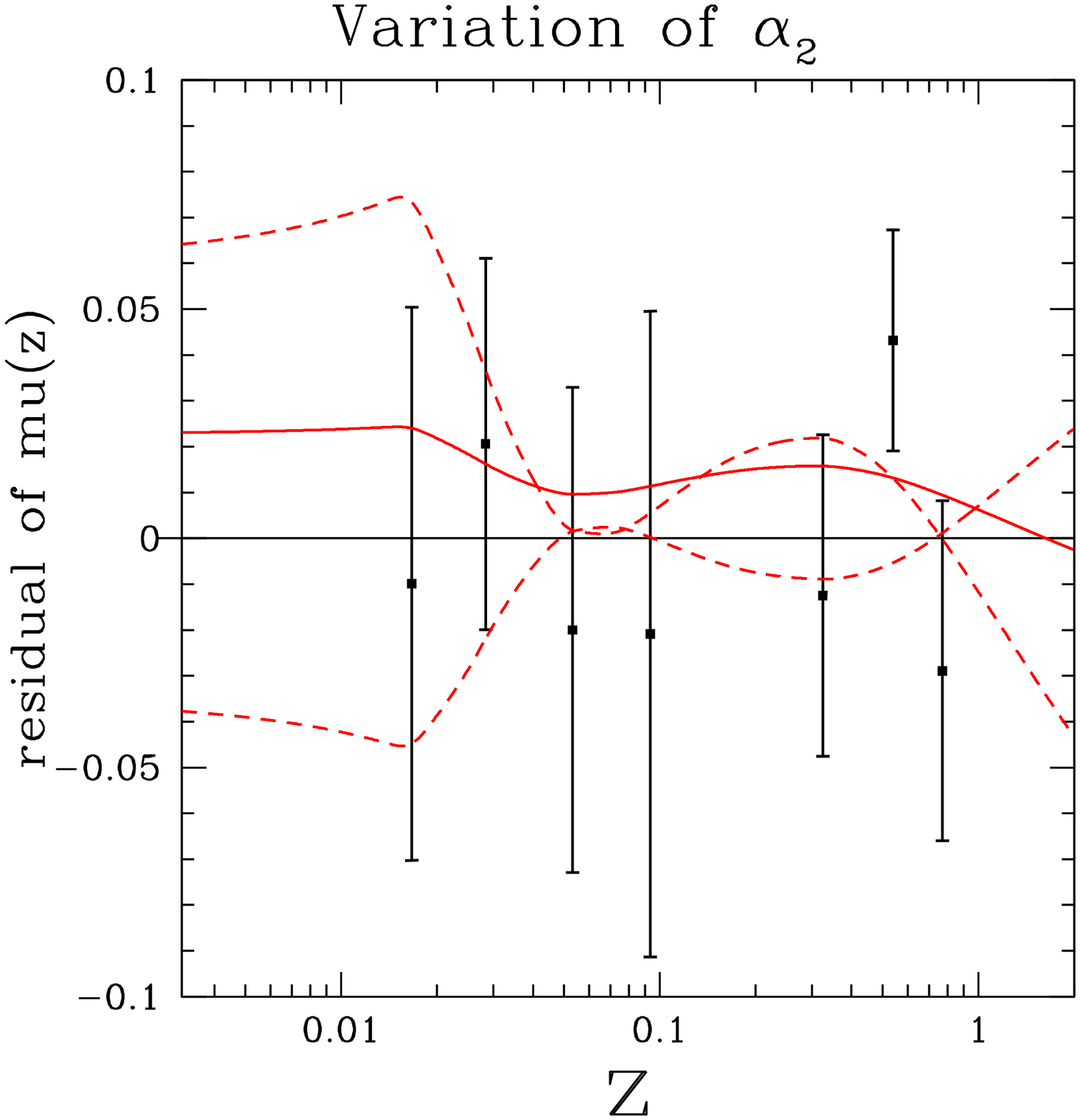}
\caption{The observable consequences of altering the mode
coefficients.  The points with error bars are the binned SNLS
residuals after subtraction of the mean $\Lambda$CDM distance modulus.
The curves are the residual mean distance moduli with different
constraints on $\alpha_i$ with $i=0$ (left panel), $i=1$ (center panel) and $i=2$
(right panel).  The constraints are that $\alpha_i$ is held fixed at 1 $\sigma$
above or 1 $\sigma$ below its mean value (dashed curves).  
As the mode number is increased, the oscillation frequency increases.}
\label{fig:onesigvar}
\end{figure*}

The interpretation of these modes is simplified by examining them in
the distance modulus space.  The effect of varying $\alpha_i$ on
distance modulus is complicated by the fact that $\alpha_i$ is
correlated with $\omega_m$, $\Omega_k$, $\alpha_0$, and the supernova
parameter $M$.  Thus in Fig. ~\ref{fig:onesigvar} we take, for
example, the parameter $\alpha_1$ in the center panel, and set it to
its mean value plus $\sigma(\alpha_1)$.  We then examine the chain in
a small region about this value to obtain the mean values for all of
the other parameters in the chain and calculate $\mu(z)$ for
this point in the parameter space.  Finally, we subtract the best-fit
$\Omega_\Lambda$ model for comparison.  We also perform the same procedure
at the mean value {\em minus} $\sigma(\alpha_1)$ and then repeat for
$\alpha_0$ and $\alpha_2$.  

Note that these modes have significant support at $z < 0.1$ and even
$z < 0.02$ and thus are affecting distances at redshifts as small as
$0.02$.  This is due to the nearby sample that spans $0.015 < z <
0.125$.  The low-redshift support in these modes makes them potentially
sensitive to large bulk flows \citep{shi97,shi98,zehavi98} that
systematically shift the redshifts of the observed supernovae from
their Hubble flow values.  \citet{hui05} find surprising sensitivity
of $w$ determinations to peculiar velocity effects in forecasts for
future observations.  Given the weight some of our best-determined
modes place on low $z$ data, peculiar velocities could potentially be
contributing significant systematic errors to our results as well.  We
address this concern below.

\begin{figure}[h]
\centerline{\scalebox{.4}{\includegraphics{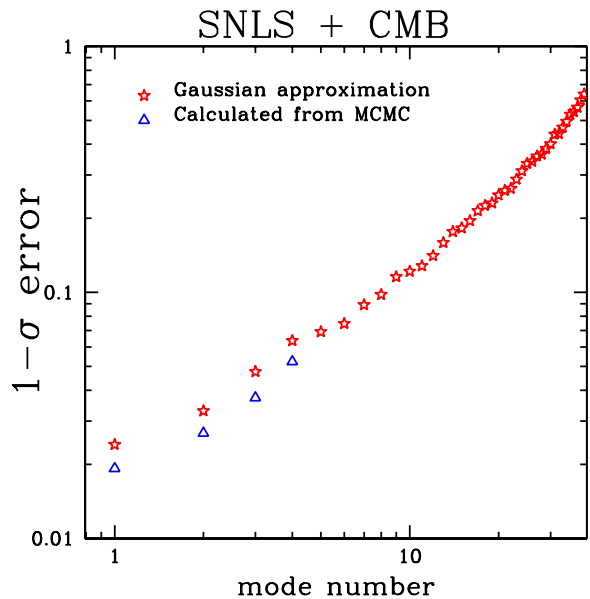}}}
\caption{The one $\sigma$ errors for the first 40 of our 50 eigenmode
amplitudes plotted for both the MCMC analysis and the Fisher matrix
analysis for SNLS + CMB.  The $1-\sigma$ errors from the Fisher matrix
are the square roots of the eigenvalues of the inverse of the Fisher
matrix associated with the respective modes.  The modes are normalized to have
length $\sqrt{N}$ where $N$ is the number of basis elements, as described
in \ref{sec:initial}.
}
\label{fig:evals}
\end{figure}

The whole spectrum of (square roots of) eigenvalues for these modes is
shown in Fig.~\ref{fig:evals}.  The stars are the result of our Fisher
matrix calculation and the triangles from MCMC.  Including more than
four modes in our MCMC calculation leads to gross degeneracies that do
not occur in the Gaussian approximation; we have not been able to
reliably calculate the spectrum beyond this first handful of modes.
For these lowest modes we see the Gaussian approximation provides a
description of the errors good to about 15\%.  For the Gold data (see
Appendix B) the Gaussian approximation is even better.

At first blush, the large number of low noise modes
contradicts what we know about the eigenvalue spectrum of $w(z)$.
\citet{knox04} found that for a {\em future space-based mission}
only a few $w(z)$ modes could be reconstructed with errors smaller than
0.1.  Here, for {\em current} data, we see eight time-varying
$\rho_x(z)/\rho_c$ modes can be reconstructed with errors smaller than
0.1.  Others have pointed out that $\rho_x(z)$ can be
reconstructed with much smaller fractional errors than $w(z)$
\citep{wang06}.  It is well known that differentiating data makes them
noisier and, roughly speaking, reconstructing $w(z)$ requires one more
differentiation of the data than a reconstruction of $\rho_x(z)$.

We only mean to point out here that there is this difference, and that
the reasons for it can be understood.  We are not using this
difference as an argument that one should consider only $\rho_x(z)$
and not $w(z)$.  Although $w(z)$ has larger error bars than
$\rho_x(z)/\rho_c$, this is not because more information is lost in a
reduction to $w(z)$.  There is a $\rho_x(z)$ for every $w(z)$,
obtainable by integration.  What happens in integrating noisy modes of
$w(z)$ to get the corresponding $\rho_x(z)$ is that the rapid
oscillations with larger departures from the fiducial value of -1 get
averaged out, suppressing the rapid oscillations.  A large amplitude
(compared to fiducial value of -1) noise fluctuation on a $w(z)$ mode
will be a much smaller fluctuation on the corresponding $\rho_x(z)$
(compared to fiducial value of $\Omega_x \rho_c$).  For discussion of
the relative merits of parameterizing dark energy by $\rho_x(z)$ and
$w(z)$ see \citet{linder04b}.

\begin{figure}[h]
\centerline{\scalebox{.4}{\includegraphics{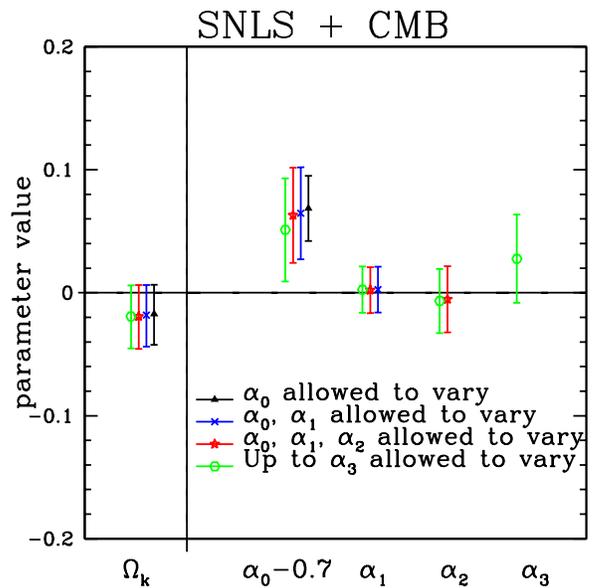}}}
\caption{Parameter estimates for the SNLS + CMB data set.  For each
parameter, we estimated the value multiple times, each time allowing a
different number of dark energy parameters to vary. } 
\label{fig:snlspvalues}
\end{figure}

In figure
\ref{fig:snlspvalues} we plot our estimates of
$\Omega_k$, $\alpha_0 - 0.7$, and some of the varying $\alpha_i$
parameters.  We subtract $0.7$ from $\alpha_0$ for plotting
convenience.  Recall that $\Omega_\Lambda = \alpha_0$
when the time-varying modes have zero amplitude.
The multiple data points for a given parameter are the results for
different numbers of parameters held fixed.

Note that the first three nonconstant mode amplitudes are consistent
with zero.  Like many others \citep{doran05,daly05,xia05,ichikawa05,
sanchez06, nesseris05}  we are finding consistency of
available supernova plus CMB data with a flat $\Lambda$CDM universe.

Note also that the curvature constraints are
robust to variation of these first few modes.  They are sufficiently
well-constrained that they do not lead to significant confusion
with the curvature.  

We also see that the amplitudes of the nonconstant eigenmodes
themselves do not change appreciably when we change how many mode
amplitudes we vary.  Within the Gaussian approximation this
independence is expected, since the modes were chosen to have
uncorrelated Fisher matrix amplitude errors.

\begin{figure}[h]
\centerline{\scalebox{.4}{\includegraphics{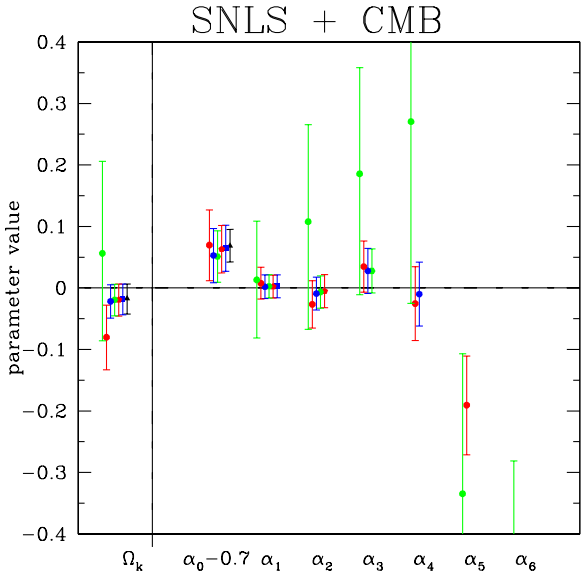}}}
\caption{As we allow more freedom in $\rho_x(z)$ by allowing more
  $\alpha_i$ to vary, the errors in the inferred cosmological
  parameters grow.  The leftmost estimate for each parameter comes
  from a chain allowing up to $\alpha_6$ to vary.  For $\alpha_0-0.7$
  this left-most estimate is off the chart.}
\label{fig:breakdown}
\end{figure}

While we have shown that for the first few dark energy parameters our
Gaussian approximation remains valid and the curvature constraint
remains robust, we expect that if we add more and more dark energy
parameters, eventually these facts will change.  We expect this for
two reasons.  First, the freedom to vary the more poorly constrainted
parameters will allow us to move far enough away in parameter space
(from the point at which the Fisher matrix was evaluated) for the
Gaussian approximation to break down.  Second, even within the
Gaussian approximation, the errors in the amplitudes of the dark energy
modes are only constructed to be uncorrelated with each other; they are
correlated with the errors in all of the other parameters, including
the displayed $\Omega_k$ and $\alpha_0$.  Thus including more modes
will weaken these constraints.  In figure \ref{fig:breakdown} we see
the results of including up to mode number 6.

The breakdown of Gaussianity is very sudden.  The moment the chain
includes the parameter $\alpha_6$, the errors on all other parameters
increase dramatically.  We believe the breakdown is driven by a
correlation between the $\alpha_6$ error and the $\alpha_0$ error,
which then allows for variations far from the fiducial model.  With
these large variations the mean values of our cosmological parameters
end up describing a cosmology that is inconsistent with other
experiments.  For example, when we include the $\alpha_6$ parameter,
$\Omega_m$ gets dragged up against the prior's upper bound of 1 so that
$\Omega_m = 0.89 +0.10 -0.27$ at 95\% confidence.  

Results from the other data set combinations are not dramatically different
to those from SNLS plus CMB that we have considered here in detail.  For
completeness, we include them in Appendix B.

We now turn to the concern that because our best-determined modes have
strong support at very low redshift they may be contaminated by
(unmodeled) peculiar velocity-induced redshifts.  We address it by
constructing modes (as described in Appendix A) that only have
variation in the region 0.2 < z < 2.0.  This region contains all the
SNLS supernovae and none of the supernovae from the nearby sample.  

\begin{figure}[b]
\centerline{\scalebox{.4}{\includegraphics{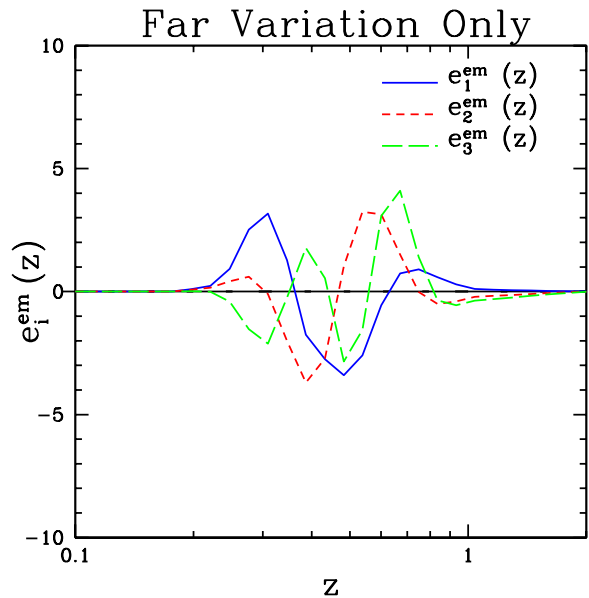}}}
\caption{The three best-measured eigenmodes for the SNLS + CMB data set
with the restriction that the modes only vary at $z > 0.2$.}
\label{fig:eigenmodes2}
\end{figure}

\begin{figure}[h]
\centerline{\scalebox{.4}{\includegraphics{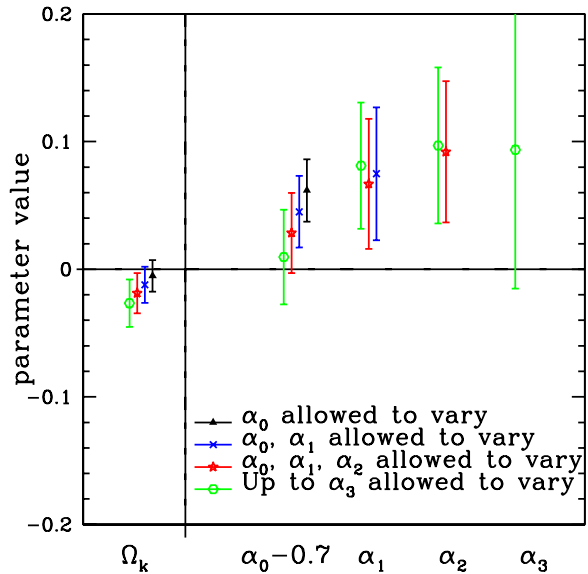}}}
\caption{Parameter estimates for the SNLS + CMB data set with the
time-varying dark energy modes as shown in Fig.~\ref{fig:eigenmodes2}.  For each
parameter, we estimated the value multiple times, each time allowing a
different number of dark energy parameters to vary. } 
\label{fig:snlspvalues2}
\end{figure}

Our best-determined modes resulting from this procedure are shown
in Fig.~\ref{fig:eigenmodes2} and the corresponding parameter 
constraints are shown in Fig.~\ref{fig:snlspvalues2}.  These 
mode amplitudes have larger error bars than the previous ones,
but are safer from contamination by peculiar velocities.  

\section{Discussion}

Our modes have been defined based on which ones can be best determined
by the data.  We now turn to consideration of the expected signal
contribution to these mode amplitudes.  

It is difficult to see how the dark energy density could vary on
time scales faster than a Hubble time.  Were the dark energy
density to start diluting with the expansion as $a^n$ we would
have $d \ln \rho/dt = n H$; i.e. the time-scale for variation would
be $1/|nH|$.  Even were the dark energy density to suddenly thermalize
as massless radiation we would only have $n=-4$.  Expressing $n$ in
terms of variation of the equation-of-state parameter away
from its value for a cosmological constant ($\delta w = w-(-1)$) we have
$n=3 \delta w$.  Thus we roughly expect $|d\ln \rho/dt| < H$.

\begin{figure}[h]
\centerline{\scalebox{.4}{\includegraphics{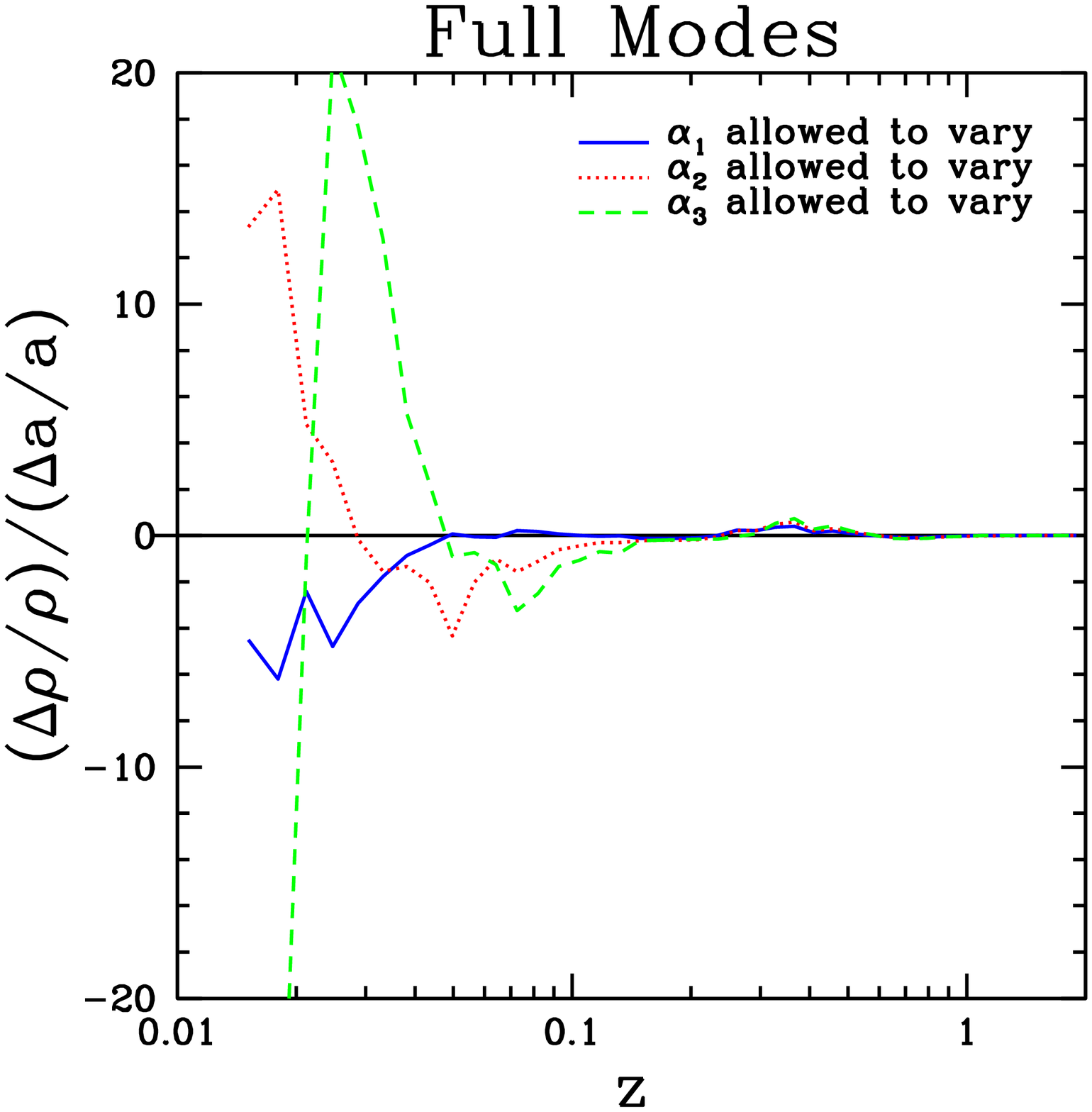}}}
\centerline{\scalebox{.4}{\includegraphics{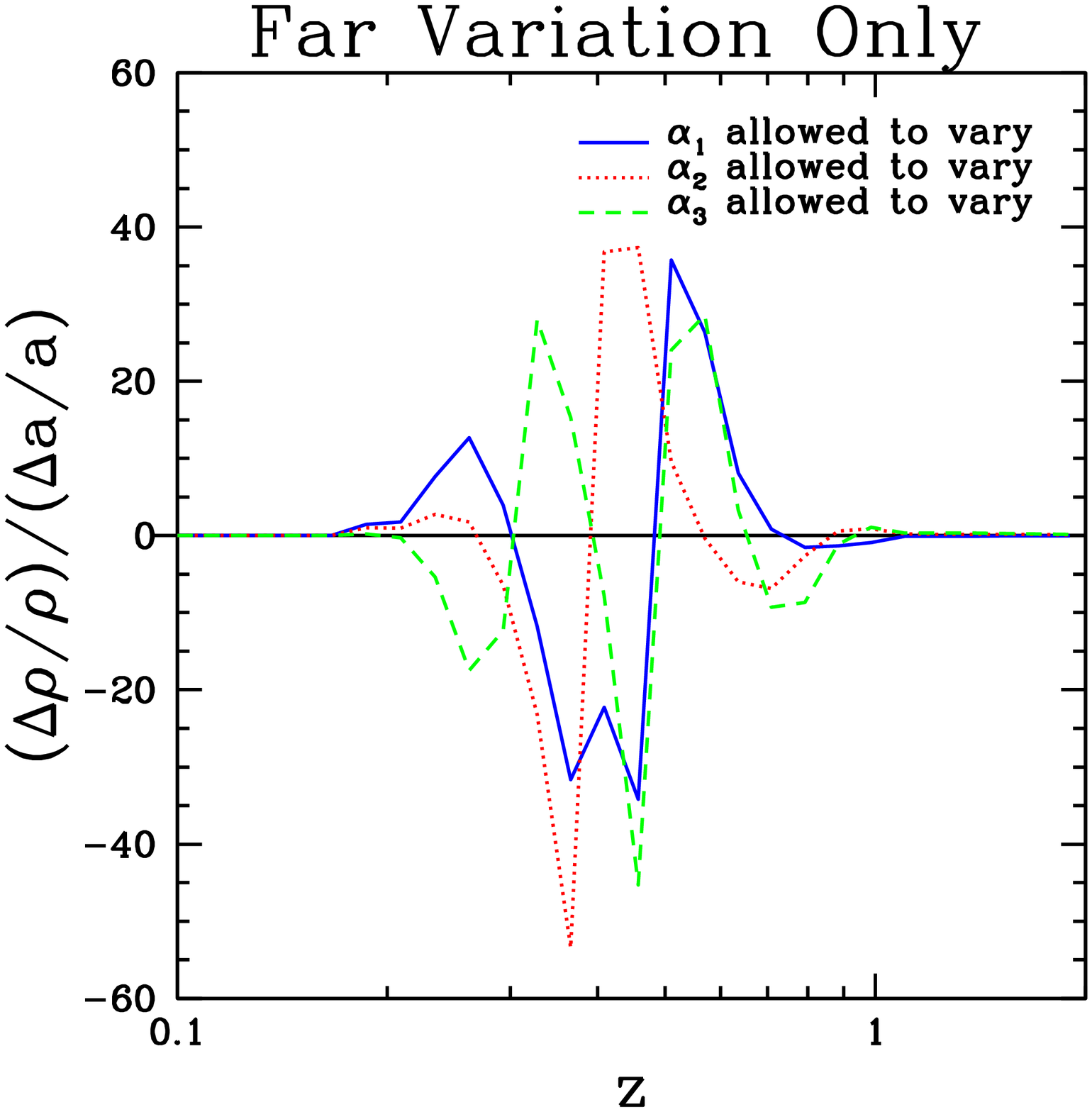}}}
\caption{$(\Delta \rho_x/\rho_x)/(\Delta a/a)$ for our modes with
amplitudes at their 68\% confidence upper limits.  This is roughly
equal to the ratio of the natural time-scale for variation ($1/H$) to
the mode variation time-scale $(d \ln \rho/dt)^{-1}$.  We see that the
mode variation time-scales are longer than the natural time scale. Top
panel: modes varying over all $z$.  Bottom panel: modes varying only
in the SNLS region.}
\label{fig:timescale}
\end{figure}

Ignoring contributions to $H$ from curvature and matter, we can write
$d\ln \rho_x/dt/H = d \rho_x/\rho_x/da/a$ which can be conveniently
calculated for our modes by finite difference.  Note that the
time-scale for variation is not a property of the mode itself, but a
property of the mode and the amplitude of the mode.  We evaluate it
for each mode with an amplitude equal to its $1 \sigma$ uncertainty.
We find they all have regions of redshift for which they exceed unity,
as can be seen in the two panels of Fig.~\ref{fig:timescale}.

Note that even though the far region modes in Fig.~\ref{fig:eigenmodes2}
look smoother than those in Fig.~\ref{fig:eigenmodes}, their amplitude 
constraints are weaker with the result that
the density variation time scales, evaluated with an amplitude of $1\sigma$,
are actually smaller than those for the full modes.  

We have also explored the idea of finding modes that maximize signal-to-noise
ratio rather than the ones that merely minimize noise.  We have
attempted to do so using the technique of signal-to-noise eigenmodes
\cite{bond95,bunn95,kl}, also known as the Karhunen-Loeve technique.

The procedure in general is for data with both a noise covariance matrix
and a signal covariance matrix; i.e., data that is modeled as
\be
\Delta_i = s_i + n_i
\ee
where the signal contribution has covariance matrix 
\be
S_{ij} = \langle \left(s_i - \langle s_i \rangle\right)  
\left(s_j - \langle s_j \rangle \right) \rangle
\ee
and similarly for the noise, $N_{ij} \equiv \langle n_i n_j\rangle$.

The procedure starts with a whitening of the data so that all modes
have errors that are uncorrelated with unit variance.  The signal
matrix in this new space is $N^{-1/2} S N^{-1/2}$.  Eigenmodes of this
`signal-to-noise matrix' have eigenvalues that are equal to the
signal-to-noise ratio which is exactly what we want.

For our application we view the density in redshift bins as the
`data'.  The trick is identifying a useful form of the signal matrix.
To date we have not found a useful form.  For the signal matrices
which we have tried, the result has been that though the modes
themselves become smoother, the error estimates increase, such that
the density variation time scale, evaluated with an amplitude of $1\sigma$,
does not decrease substantially.

\section{Conclusions}

The single most important question about the dark energy is
whether it is a cosmological constant.  We have presented
a method to test this hypothesis by looking for the deviations
from constant density that are easiest to detect.  The best-determined
time-varying modes of $\rho_x(z)$ are found by use of a Gaussian 
approximation to the parameter likelihood, but the estimates
of the amplitudes of these modes do {\em not} use the Gaussian
approximation.  We have applied the method to supernova, CMB and BAO data.  

We found that SNLS + CMB data are capable of constraining the
amplitudes of three time-varying modes of $\rho_x(z)$ to better than
5\%.  None of them are significantly different from zero.  We found
that if only these well-constrained modes are allowed to vary then the
curvature is also not significantly different from zero.  We found
that the best-determined mode amplitudes have highly independent
errors, which within the Gaussian approximation occurs by design.  We
found that they are sensitive to density variations at very low ($z
\sim 0.02$) redshifts.  We noted that these redshifts are sufficiently
small to raise concerns about the effects of bulk flows.  We pointed
out that the best-determined modes are useful not only for detection of
time variation, but also for ferreting out systematic errors by studying
how the amplitude estimates fluctuate with varying data subsample.  

We pointed out that, by quite general considerations of the
time-scales expected for dark energy density variation, at the
achieved sensitivity levels we do not expect a detectable signal.
Because we strongly expect a null result, our application can be
viewed as a consistency test which has the potential for exposing
systematic errors.  This situation is analogous to that with the
(theoretially unexpected) B-modes which are a useful diagnostic in
cosmic shear observations \citep[e.g.][]{crittenden02}.  As the 
statistical constraining
power of the data improves we will begin to probe the regime where
signals can be expected.  We also suggested
that with the adoption of an appropriate signal covariance matrix we
could use the signal-to-noise eigenmode technique to maximize
signal-to-noise ratio rather than simply minimizing noise.  

In Appendix B our results show that the BAO data add significant
additional constraining power to our cosmological parameter estimates.
Unfortunately, the convenient reduction to a single number, though
valuable for other applications where the dark energy is assumed to
have a constant equation-of-state parameter, is not clearly applicable
to our problem.  It would be useful to have a reduction of the BAO
data to constraints on the distances to several redshifts, or as a
single constraint on some linear combination of redshifts.

\acknowledgments We thank D. Huterer for a useful conversation that
stimulated this work as well as A. Albrecht and M. Hudson for useful
conversations.  This work was supported in part by NASA grant
NAG5-11098 and NSF grant 0307961.

\bibliography{./cmb4}

\appendix

\section{Pre-diagonalization modes}

Here we describe our construction of the initial basis that we use
as the first step in creating the time-varying modes with uncorrelated
amplitude errors.  

The first basis vector is a constant, which before normalization looks
like (1, 1, 1, ... , 1).  The next $\frac{N}{2}$ basis vectors take the
form (0, 0, 1, -1, 0, ..., 0).  As long as none of the non-zero values
in this second set of basis vectors overlap with one another, they
will be orthogonal to one another, and each is clearly orthogonal to
the constant vector.

To fill out the space further with a third set of basis vectors we
notice that each of the second set of basis vectors is orthogonal to
the constant vector, but only in steps of two.  A new basis vector
that maintains the same value in steps of two will be orthogonal to
each previous basis vector: (1, 1, -1, -1, 0, ..., 0).  To find the
next one we consider that this basis vector is orthogonal to (1, 1, 1,
1, 0, ..., 0), so with the previous argument all the basis vectors
made so far will clearly be orthogonal to (1, 1, 1, 1, -2, -2, 0, ...,
0).  The next basis vector will then include 6 1's and two -3's, and
so on.  Careful counting reveals that there are $\frac{N}{2} - 1$ of
this third set of basis vectors, which will complete our basis if we
have an even number of elements.  If we have an odd number of elements
we can add a final basis vector that takes the form (1, 1, 1, ..., 1,
-(n-1)), since every element of the second set of basis vectors
contains a zero in the last position.

Below is an example $9 \times 9$ basis (with vectors as rows), 
before normalization.

\begin{center}
\be\begin{array}{rrrrrrrrr}
1 & 1 & 1 & 1 & 1 & 1 & 1 & 1 & 1\\
1 &-1 & 0 & 0 & 0 & 0 & 0 & 0 & 0\\
0 & 0 & 1 &-1 & 0 & 0 & 0 & 0 & 0\\
0 & 0 & 0 & 0 & 1 &-1 & 0 & 0 & 0\\
0 & 0 & 0 & 0 & 0 & 0 & 1 &-1 & 0\\
1 & 1 &-1 &-1 & 0 & 0 & 0 & 0 & 0\\
1 & 1 & 1 & 1 &-2 &-2 & 0 & 0 & 0\\
1 & 1 & 1 & 1 & 1 & 1 &-3 &-3 & 0\\
1 & 1 & 1 & 1 & 1 & 1 & 1 & 1 &-8\\
\end{array}\ee
\end{center}

Each of these basis vectors defines an $e^0_i(z)$.  To
do this, we connect each element in the vector with a particular
position in $z$.  The value of $e^0_i(z)$ between these positions is
given by linear interpolation.  These basis vectors $e^0_i(z)$ are
used to define $\rho_x(z)$ as in equation (\ref{eq:rhox}):
\be
\label{eq:rhox}
\rho_x(z) = \rho_c(z=0)\sum_{i=0}^{N}\alpha^0_i e^0_i(z).
\ee

It is worth noting that while we demanded that the non-constant
eigenmodes were also zero-mean, there is an arbitrariness to the
meaning of zero mean.  It depends on the definition of the dot 
product, which is, in turn, basis-dependent.  For example,
the average value of a mode sampled in $\ln(z)$ is different 
from the average value of the same mode sampled in $z$.    

Given this arbitrariness, it is worth considering relaxing the 
zero-mean condition.  With that condition abandoned, it becomes possible to
use the new freedom to decorrelate the time-varying mode amplitude
errors with the constant mode amplitude error.  In principle one could
do so in two steps: first add the right amount of constant to each of
our time-varying modes to decorrelate it with the constant.  Then
re-diagonalize in the time-varying subspace.  In practice we have been
unable to circumvent numerical instabilities that occur when we
attempt this procedure.

In order to suppress variation of the near region, we chose to modify our
pre-diagonalization basis.  We select a basis where the first mode is a
constant. The second is constant over the near region and the far region, with
the two differring in sign, with the values of the near and far regions chosen
to enforce orthogonality with the constant mode.  The next group of modes
varies only in the near region, with the far region set to zero.  Its structure
is the same as that described earlier for the varying basis elements.  The
final group of modes again has the same structure, but varies only in the far
region.  An example $9 \times 9$ basis is shown below (before normalization).
                                                                                   
\begin{center}
\be\begin{array}{rrrrrrrrr}
1 & 1 & 1 & 1 & 1 & 1 & 1 & 1 & 1\\
5 & 5 & 5 & 5 & -4 & -4 & -4 & -4 & -4\\
1 & -1 & 0 & 0 & 0 & 0 & 0 & 0 & 0\\
0 & 0 & 1 & -1 & 0 & 0 & 0 & 0 & 0\\
1 & 1 & -1 & -1 & 0 & 0 & 0 & 0 & 0\\
0 & 0 & 0 & 0 & 1 & -1 & 0 & 0 & 0\\
0 & 0 & 0 & 0 & 0 & 0 & 1 & -1 & 0\\
0 & 0 & 0 & 0 & 1 & 1 & -1 & -1 & 0\\
0 & 0 & 0 & 0 & 1 & 1 & 1 & 1 & -4\\
\end{array}\ee
\end{center}

The location of the split is chosen to coincide with the desired location
before which no variation will occur.  Once this basis is chosen, we carry out
the diagonalization procedure as described in \ref{sec:initial}, with the
exception that we choose the subset of the covariance matrix that corresponds
to those basis elements which only vary in the far region.  

\section{Other Data Combinations}

\begin{figure*}                                                                
\plotthree{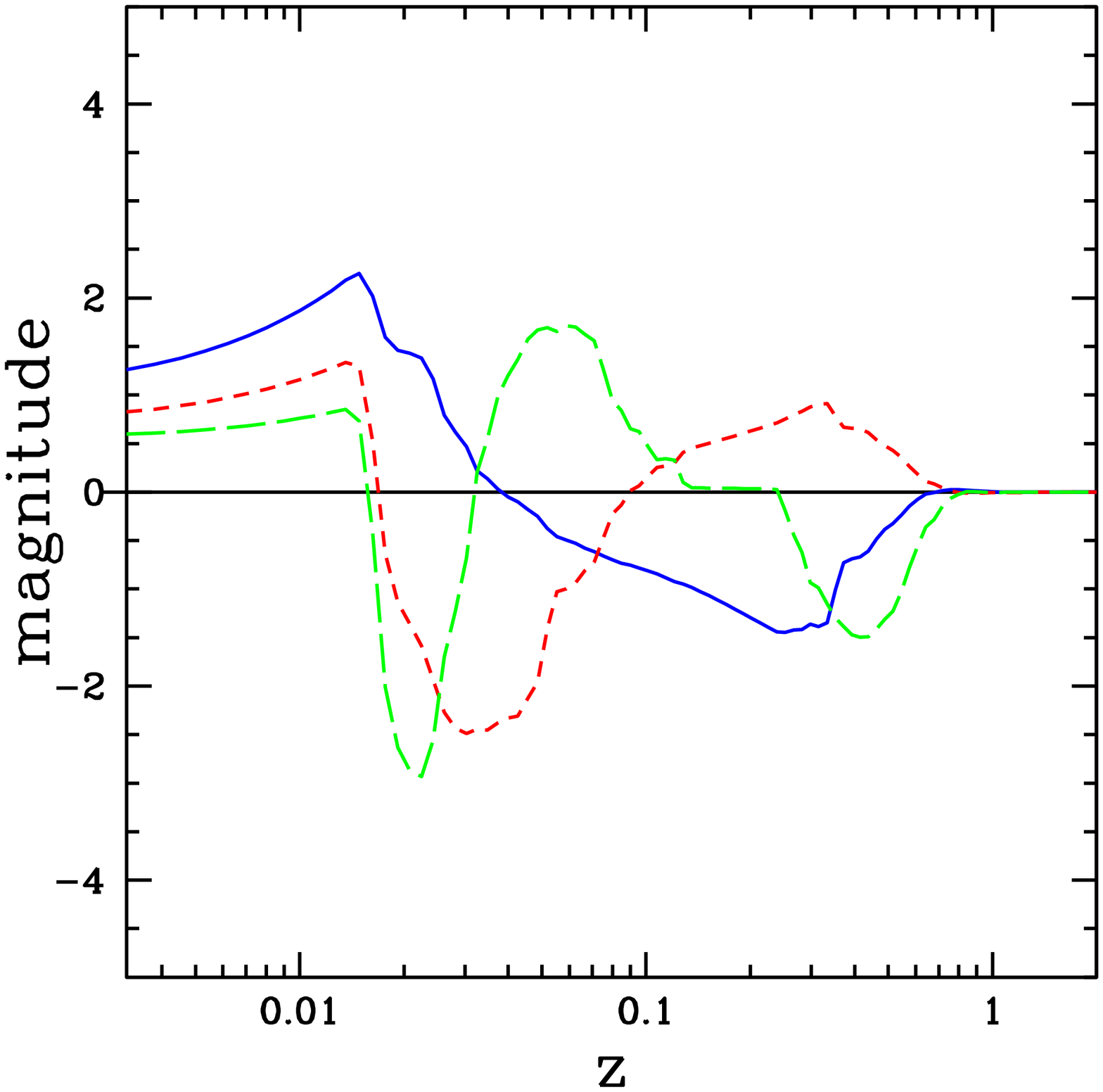}{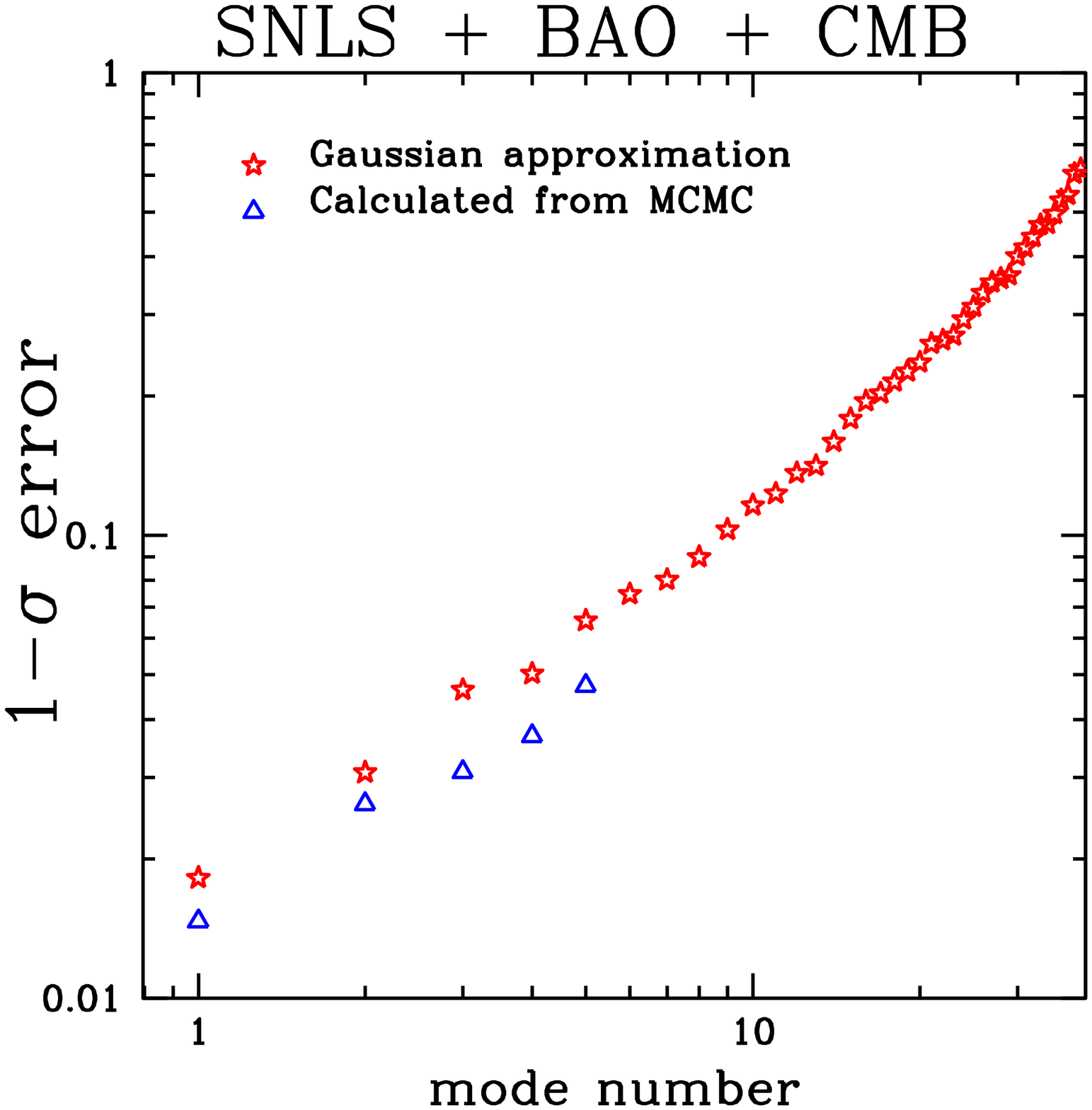}{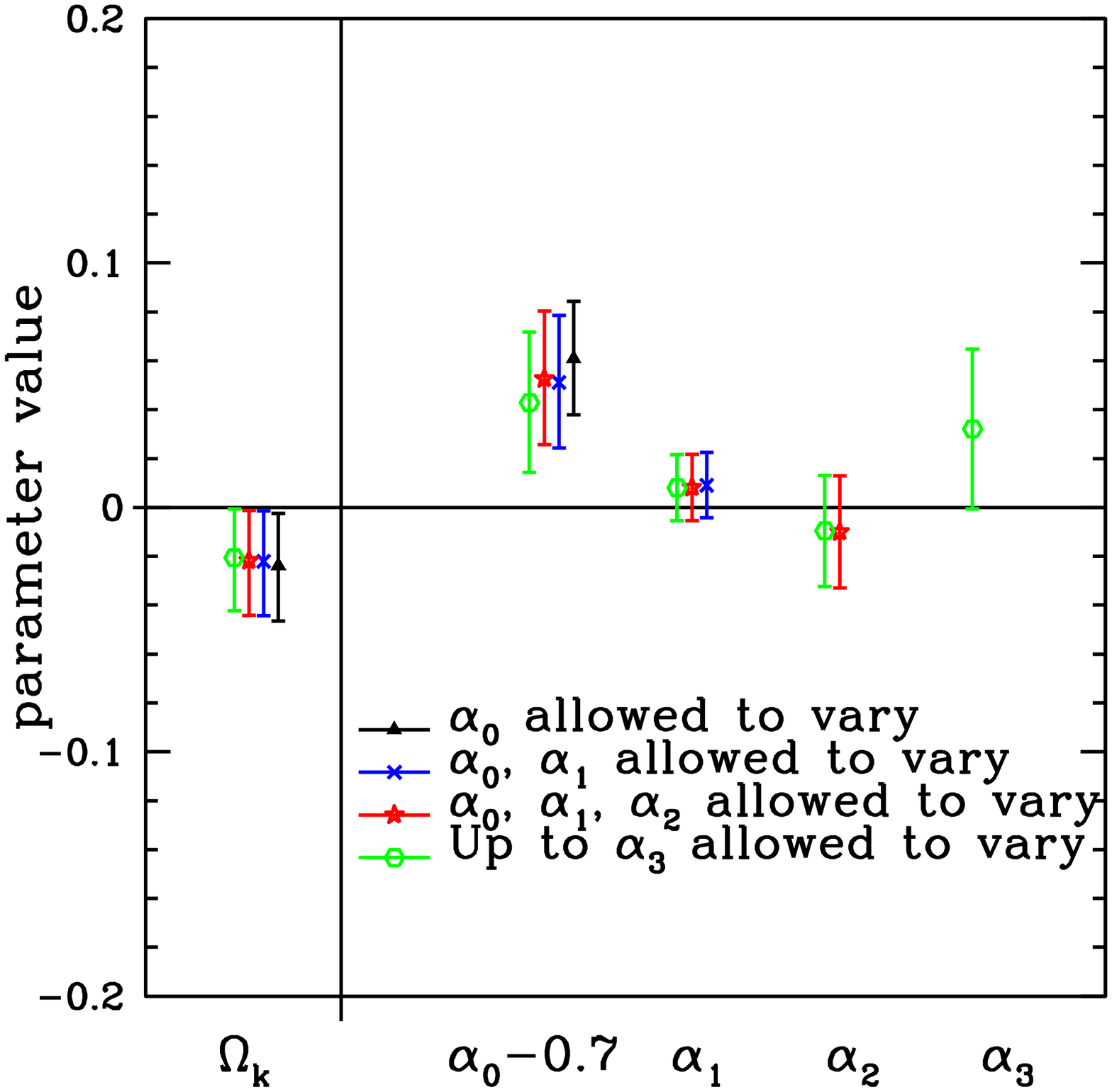}
\caption{Eigenmodes (left panel), Eigenvalue spectrum (center panel) and
parameter estimates (right panel) for BAO plus SNLS plus CMB data. }
\label{fig:snlsbaocmb}
\end{figure*}

\begin{figure*}
\plotthree{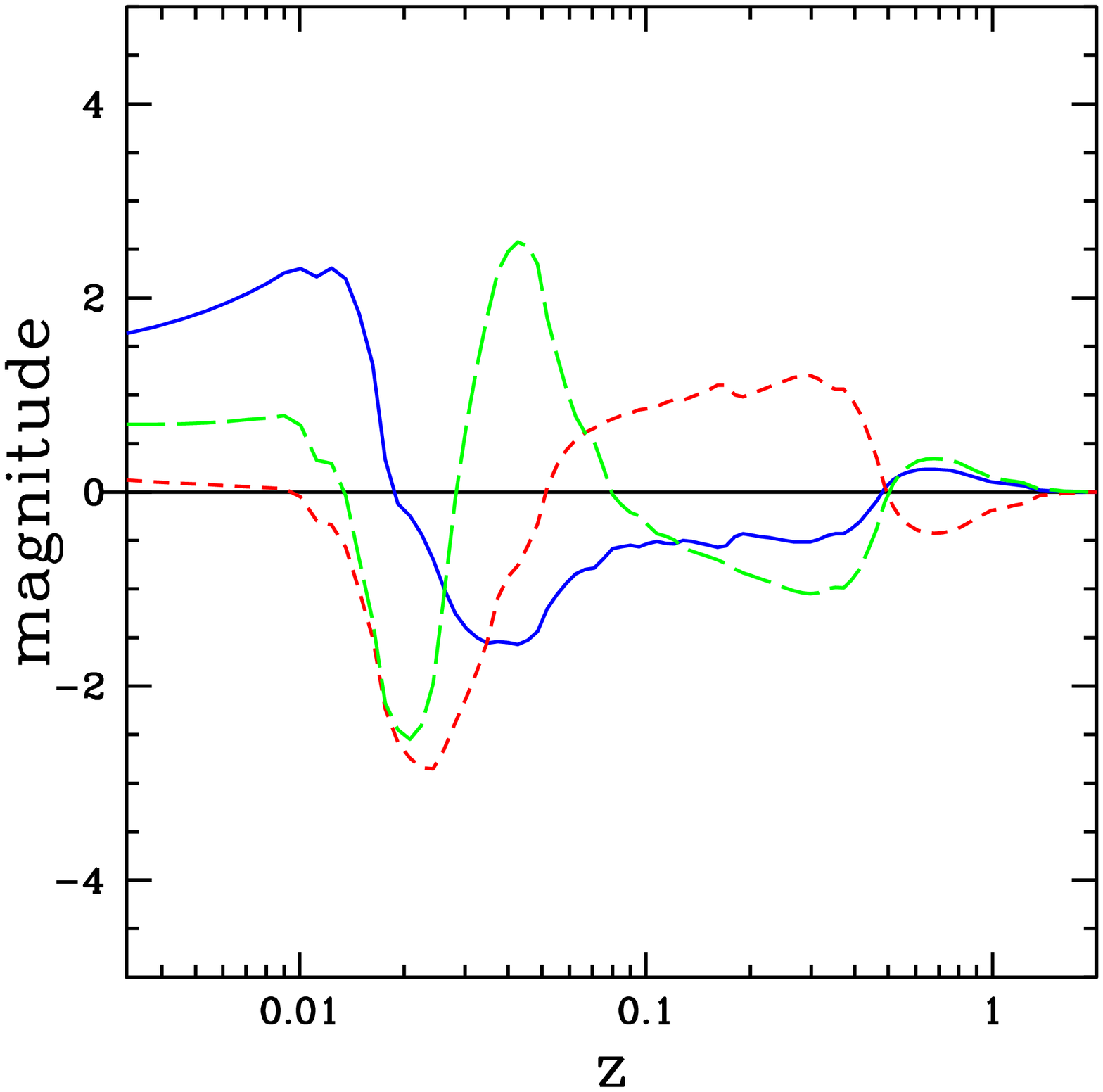}{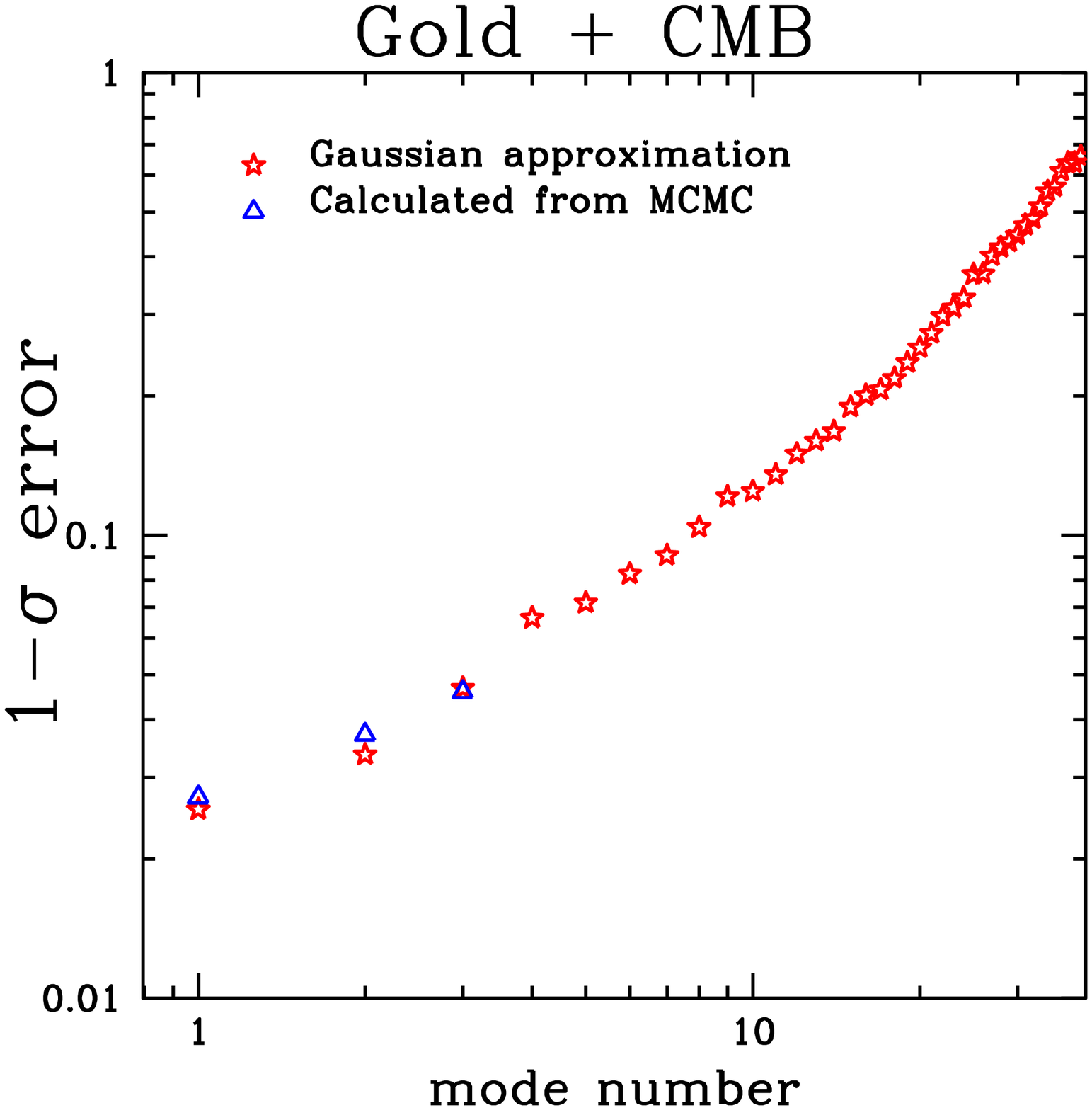}{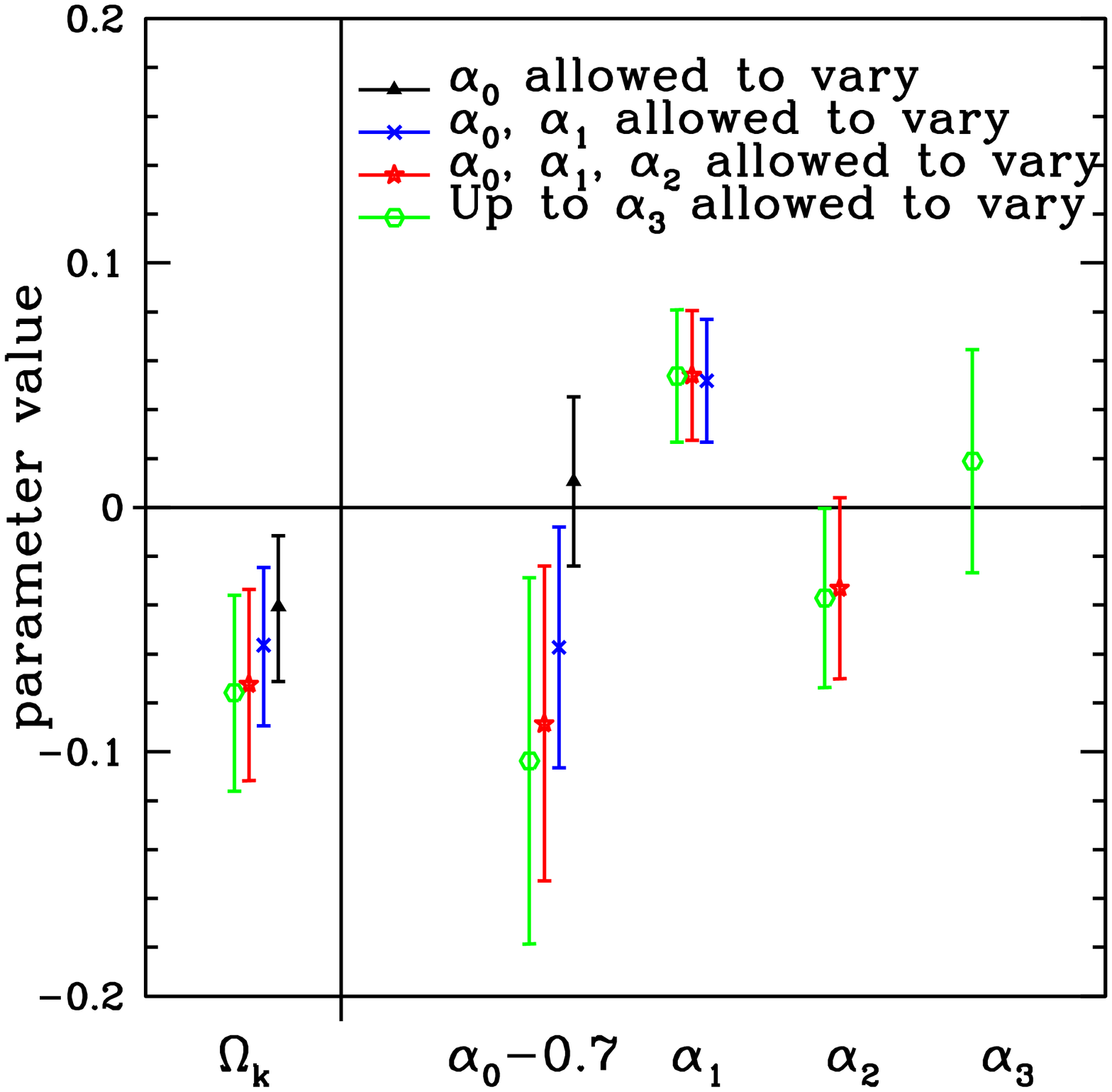}
\caption{As above, but for the Gold supernova data set combined with
the CMB.}
\label{fig:riesscmb}
\end{figure*}

\begin{figure*}
\plotthree{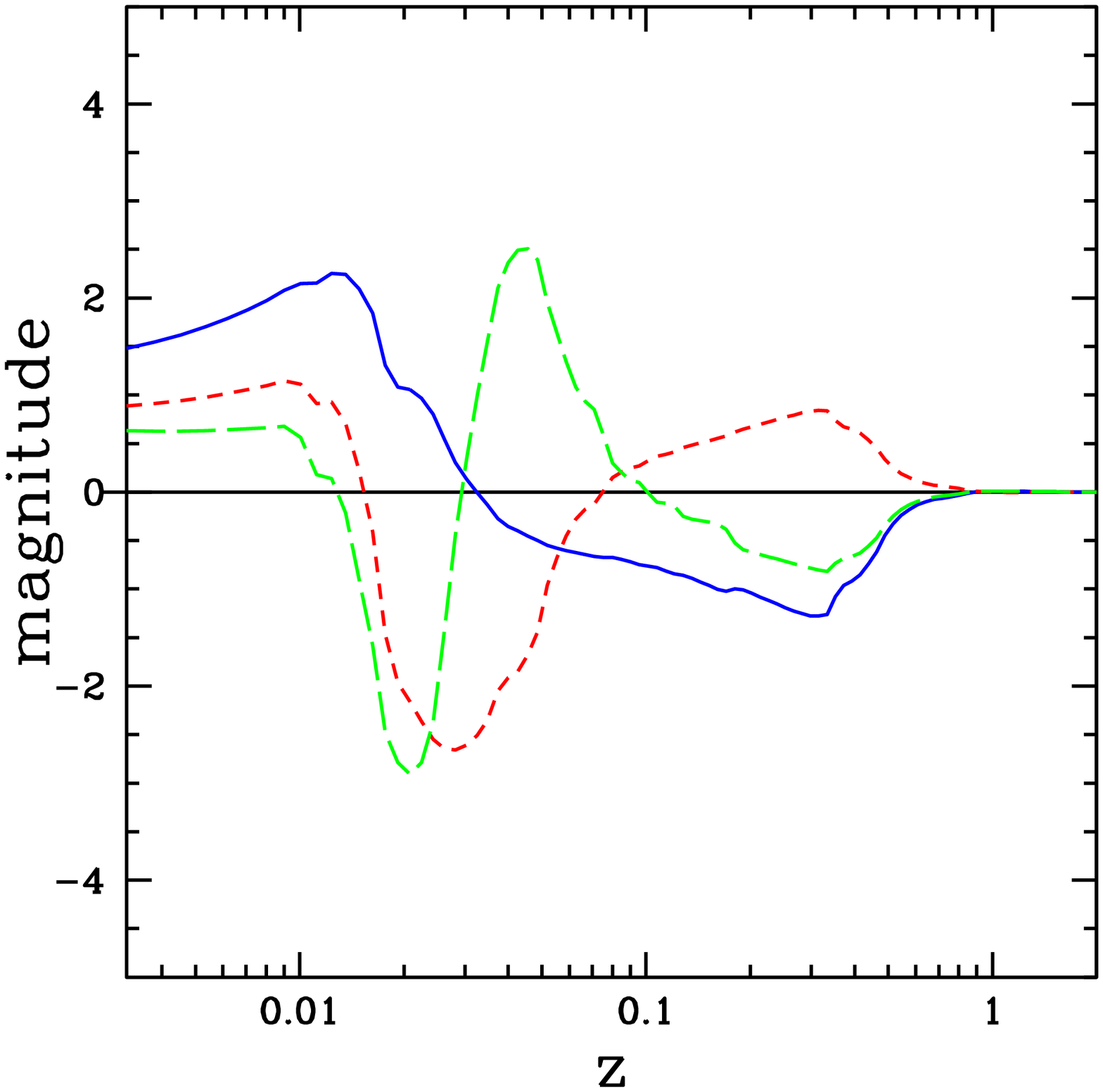}{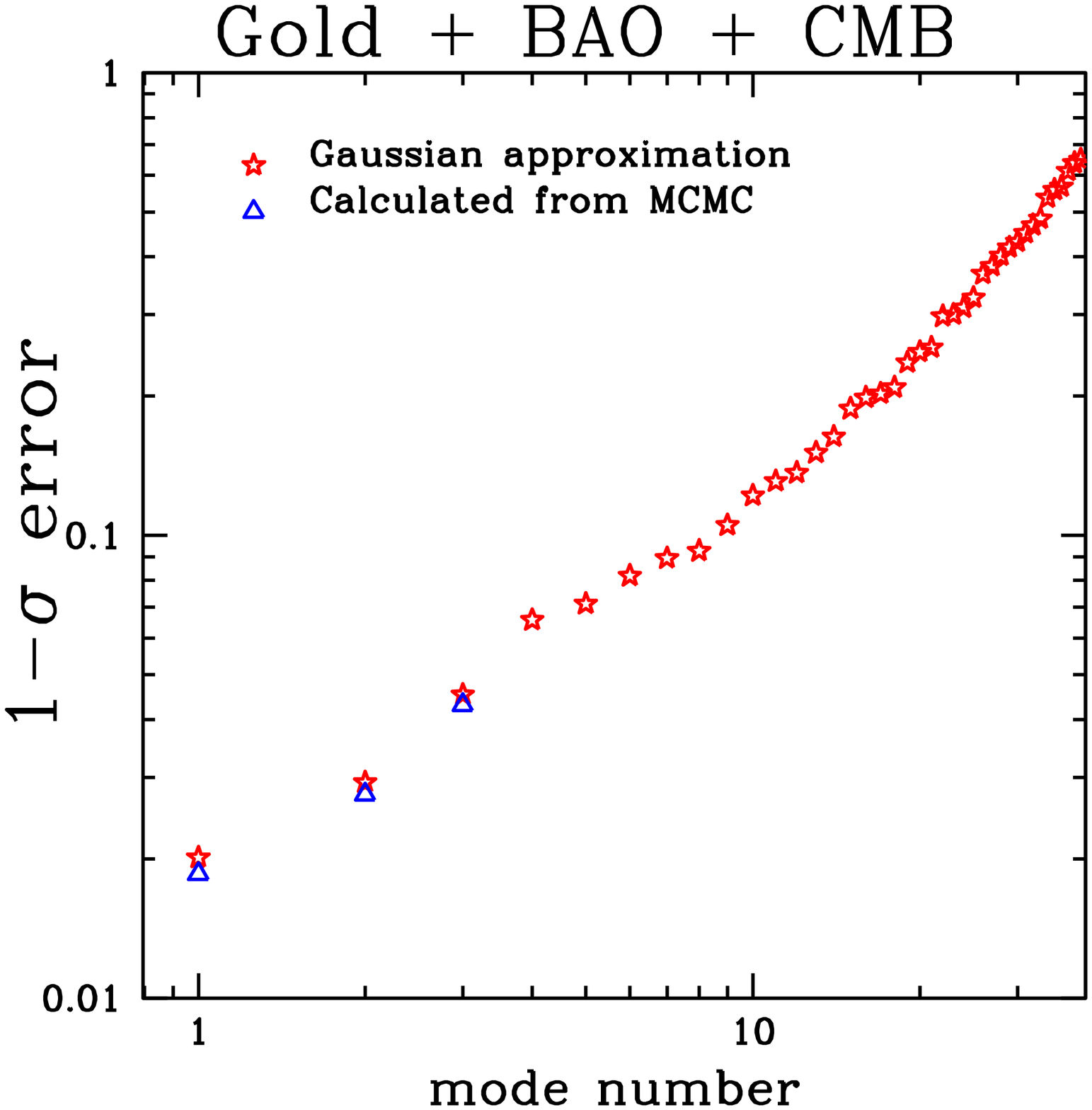}{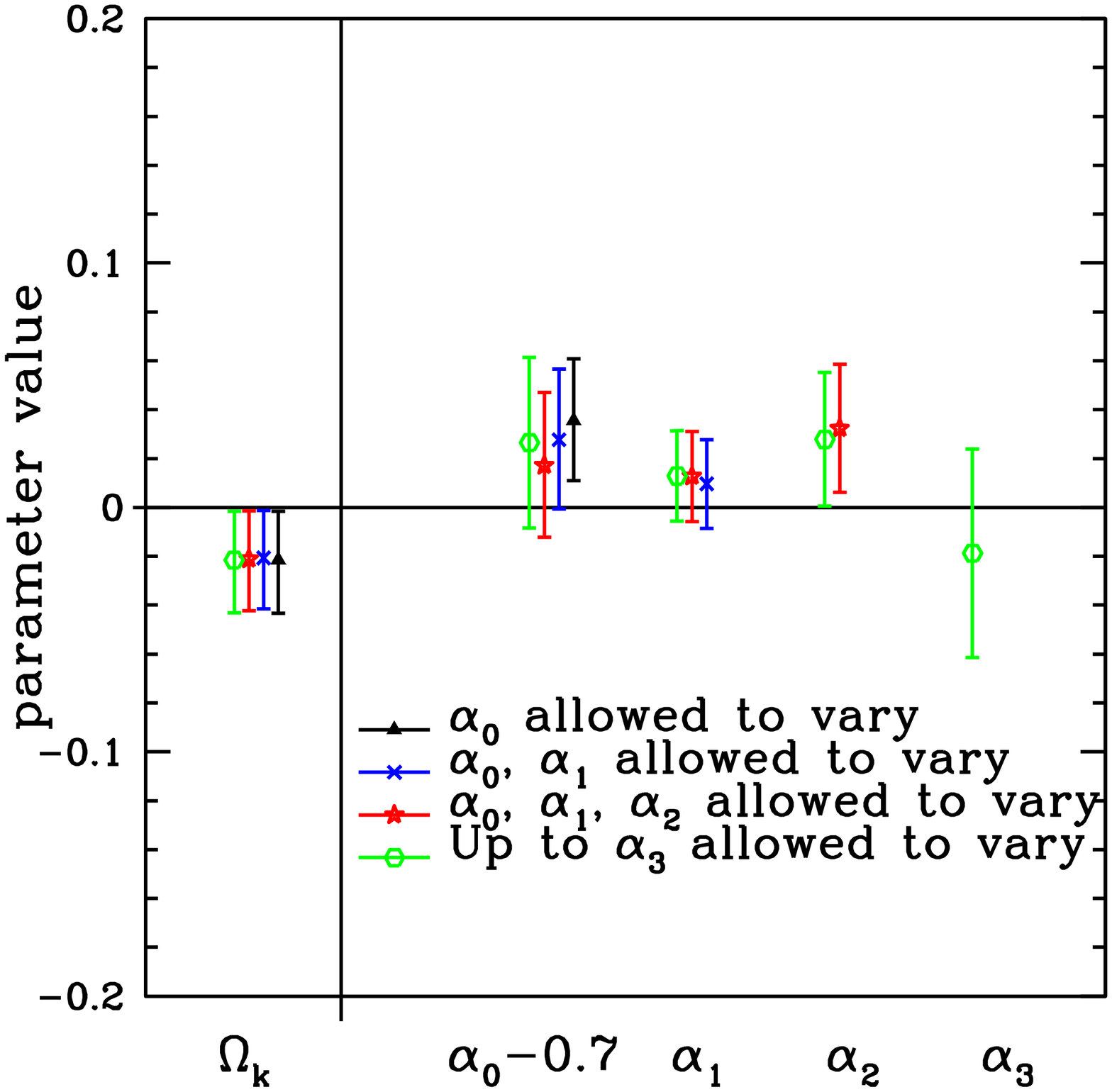}
\caption{As above, but for the Gold supernova data set combined with
BAO plus CMB.}
\label{fig:riessbaocmb}
\end{figure*}

Here we show results from application to other data combinations and
comment on the difficulty of including the BAO data.  The results from the various other data set combinations we calculated are shown in figures \ref{fig:snlsbaocmb}, \ref{fig:riesscmb} and \ref{fig:riessbaocmb}.

Adding in the BAO data, we find that we gain a fair amount of extra
constraining power for both curvature and our dark energy
parameters.  For the SNLS data adding in the BAO data gives us the
ability to measure one more dark energy parameter.  Interpretation of 
these results though is severely hampered by the manner in which we 
incorporated the BAO data.  As described in the text we utilized the
\citet{eisenstein05} reduction of their data to a constraint on a combination
of $D_M(z)$ and $H(z)$ at $z=0.35$ and the matter density (their $A$ parameter).
Although they tested the validity of this data compression for constant
$w$ models, we are not sure how valid it is for the more general dark-energy
model space that we consider.  We include the BAO data only to show the
potential statistical power of that data for an analysis such as ours.  

Artifacts of the compression to $A$ could be seen in the modes had we
not employed the following trick:  when calculating the Fisher matrix
in order to get the modes, we held the $H(z)$ in $A$ fixed.  If we
were to include $\partial H(z) / \partial \alpha_i$ as a contribution
to $\partial A/\partial \alpha_i$ then there would be a large spike
in the modes at $z=0.35$.  This change to the eigenmodes makes no
measurable difference to the parameter estimates.

\end{document}